\documentclass[aps,prb,showpacs,twocolumn]{revtex4-1}

\usepackage{amsmath,graphics}
\usepackage[next]{inputenc}
\usepackage[dvips]{epsfig}
\usepackage[colorlinks=true,citecolor=blue,linkcolor=blue]{hyperref}
\usepackage{bbm}
\usepackage{bbold}

\def\be{\begin{equation}}
\def\ee{\end{equation}}

\def\bi{\begin{itemize}}
\def\ei{\end{itemize}}
\def\bn{\begin{enumerate}}
\def\en{\end{enumerate}}
\def\bea{\begin{eqnarray}}
\def\eea{\end{eqnarray}}

\def\ba{\begin{array}}
\def\ea{\end{array}}
\def\bd{\begin{displaymath}}
\def\ed{\end{displaymath}}

\begin{document}
\title{Multi-orbital Effects on Thermoelectric Properties of Strongly Correlated Materials}
\author{Mehdi Kargarian}
\email{kargarian@ph.utexas.edu}
\affiliation{Department of Physics, The University of Texas at Austin, Austin, TX 78712, USA}
\author{Gregory A. Fiete}
\affiliation{Department of Physics, The University of Texas at Austin, Austin, TX 78712, USA}

\begin{abstract}
The effects of electronic correlations and orbital degeneracy on thermoelectric properties are studied within the context of multi-orbital Hubbard models on different lattices. We use dynamical mean field theory with a modified version of iterative perturbation theory as a solver to calculate the self-energy of the models in wide range of interaction strengths. The Seebeck coefficient, which measures the voltage drop in response to a temperature gradient across the system, shows a non-monotonic behavior with temperature in the presence of strong correlations. This anomalous behavior is associated with a crossover from a Fermi liquid metal at low temperatures to a bad metal with incoherent excitations at high temperatures, and is qualitatively captured by the Kelvin formula but not quantitavely. We find that for interactions comparable to the bandwidth the Seebeck coefficient acquires large values at low temperatures. Moreover, for strongly correlated cases, where the interaction is larger than the band width, the figure of merit is enhanced over a wide range of temperatures because of decreasing electronic contributions to the thermal conductivity.  We also find that multi-orbital systems will typically yield larger thermopower and figure-of-merit compared to single orbital models over a temperature range possibly relevant to applications.                
\end{abstract}
\date{\today}


\maketitle
\section{Introduction \label{intro}}
The thermoelectric power of materials is a measure of the voltage drop in response to a temperature gradient across the system, so thermoelectric materials can be used to convert heat to electricity, or can alternatively be used for refrigeration.\cite{Mahan:phyicstoday97, Snyder:nmat08}  In order to design efficient thermoelectric materials a variety of conflicting properties, such as large electrical conductivity, small thermal conductivity, and large thermopower should be optimized to obtain a large figure-of-merit, $ZT$, where  $ZT=\frac{S^2\sigma}{\kappa}T$, and $S$, $\sigma$ and $\kappa$ are the Seebeck coefficient (thermopower), electrical conductivity, and thermal conductivity, respectively, and $T$ denotes the absolute temperature. The thermal conductivity is typically composed of two parts: electronic and lattice contributions as $\kappa=\kappa_e+\kappa_l$. As the relation for $ZT$ suggests, reducing the lattice contribution, $\kappa_l$,  can enhance the thermoelectric properties. This can be accomplished by nanostructural engineering, which leads to enhanced scattering of phonons at interfaces or by atomic disorder.\cite{Snyder:nmat08, Poudel:Sceince2008}

A second route to increase the figure-of-merit would be to increase the power factor $S^2\sigma$, {\it i.e} the numerator of $ZT$, by increasing the thermopower $S$. Since both electrons and holes contribute to carrier transport induced by a temperature gradient but with opposite sign,  heavily doped semiconductors (which typically have a large particle-hole asymmetry) yield a large thermopower.  In contrast, metals with approximate particle-hole symmetry have vanishingly small thermopower. For a single-band model, increasing the effective mass of carriers enhances the thermopower, as for the nearly free electron gas the Seebeck coefficient is given by\cite{Cutler:prb1964}  $S=(\pi^2k_{B}^2T/3e)(8m^{*}/h^2)(\pi/3n)^{2/3}$, which immediately implies that strong correlations in narrow bands (which would increase the effective mass) lead to a large thermopower.\cite{Chaikin:prb1976, Palsson:prl1998, Oudovenko:prb2006, Tomczak:prb10, Mukerjee:APL2007} This is indeed the cause for the observed large Seebeck coefficients in transition metal compounds, such as $\mathrm{FeSi}$,\cite{Sakai:JPSP07} $\mathrm{Na_xCoO_2}$\cite{Terasaki:prb97,Haerter:prl06}, $\mathrm{Sr_xLa_{1-x}TiO_3}$\cite{Okuda:prb2001} and $\mathrm{FeSb_2}$\cite{Bentien:EPL2007}, with promising values of Seebeck coefficients $S\mathrm{\sim 100~ \mu V/K}$ at 300 K for $\mathrm{Na_xCoO_2}$ and a colossal value $S\mathrm{\sim 45~  mV/K}$ at 12 K for $\mathrm{FeSb_2}$. These materials are strongly correlated with an interaction energy scale given by the Hubbard interaction $U$, whose magnitude is usually of the order of or larger than the bandwidth. Hence, it is important to understand which correlated materials may have a large Seebeck coefficient and power factor. Of course, most oxides also have a large thermal conductivity, but the recent developments in growing thin films and engineering heterostructures can significantly reduce the phonon thermal conductivity.\cite{Venkatasubramanian:na2001,Ravichandran:prb12}                      

In this work, we revisit the effect of electronic correlations on the thermoelectric properties of solid state systems. In particular, we consider the effect of orbital degeneracy on the thermoelectric properties of strongly correlated systems on different lattices. Lattices with different crystallographic structures will yield different degrees of asymmetry in the density of states (DOS), which will influence the thermopower. A number of previous works have considered particular limiting cases of thermopower such as the atomic limit and/or high temperatures.\cite{Koshibae:prb2000, Mukerjee:prb2005,Uchida:prb11,Okuda:prb2001,Oguri:prb1990}  Here, we will focus on the interesting regime where correlations are of the same order of or larger than the bandwidth over the entire range of temperatures.  Earlier works have discussed the degeneracy of magnetic configurations and its role in the large thermopower of $\mathrm{Na_xCoO_2}$,\cite{Koshibae:prb2000, Mukerjee:prb2005} and $\mathrm{Sr_xLa_{1-x}TiO_3}$.\cite{Okuda:prb2001}   The high temperature limit in $\mathrm{La_{1-x}Sr_{x}VO_{3}}$ has been studied in Ref.[\onlinecite{Uchida:prb11}], and single band\cite{Paul:prb2003} and atomic limits\cite{Oguri:prb1990} of the Hubbard model have been considered.  Besides the thermoelectric properties, many other physical phenomena such as colossal magnetoresistance and triplet superconductivity in $\mathrm{La_{1-x}Sr_{x}MnO_{3}}$\cite{Tokura:JPSJ94} and $\mathrm{Sr_2RuO_4}$,\cite{Maeno:na1994} respectively, and superconductivity in iron-based compound $\mathrm{LaFeAsO_{1-x}F_x}$\cite{Kamihara:JACS2008} are known to be associated with correlated multi-orbital physics. 

In this work, we show that the orbital degeneracy can increase the thermopower over a wide range of temperatures possibly relevant to applications. In the models we study here, we assume the magnetic correlations are sufficiently weak that the magnetism is not the driving force behind the metal to insulator transition. Instead, we assume the Hubbard interaction drives a non-magnetic Mott insulating state.  With this assumption, the properties of these models can be calculated by use of dynamical mean field theory (DMFT) as detailed in the subsequent sections.      

The remainder of the paper is organized as follows.  In Sec.\ref{model} we describe the multi-band Hubbard model we use in our study and detail the relevant aspects of dynamical mean-field theory that we employed.  In Sec.\ref{one_orbital} we consider a few single-orbital models that we will use as a benchmark for subsequent multi-orbital models that we present in Sec.\ref{two_orbital}.  Finally, in Sec.\ref{conclusions} we give our main conclusions and discuss the implications of our work for real materials. 

\section{model and method \label{model}}
We consider a multi-orbital Hubbard model in which every site has a few degenerate orbitals. Including intra- and inter-orbital interaction between electrons, the Hamiltonian is
\bea \label{H} H&&=-\sum_{\langle ij\rangle}t_{mm'}c^{\dag}_{im\sigma}c_{jm'\sigma}+U\sum_{i}n_{im\uparrow}n_{im\downarrow} \nonumber\\ 
&&V\sum_{i,m\neq m'}n_{im\uparrow}n_{im'\downarrow}+V'\sum_{i,m\neq m'}n_{im\sigma}n_{im'\sigma},\eea where $m$ and $\sigma$ stand for orbital and spin degrees of freedom, respectively, and the summation over repeated indices are assumed.  Here, $V$ and $V'$ are given in terms of the Hubbard interaction $U$ and Hund's coupling $J$ as $V=U-2J$ and $V'=V-J$. As we also consider a single band Hubbard model in this work, the latter is simply given by $m=1$ and $V$=$V'$=$0$. 

Because we are interested in the strong interaction limit where the kinetic energy scale set by the hopping integrals $t_{mm'}$ are comparable to the Coulomb interaction energy scale or smaller, a perturbation expansion around the band limit can not describe the low energy excitations of the system properly.  A practical and capable method to handle this regime is dynamical mean-field theory (DMFT).\cite{Georges:rmp1996} Taking the limit of large coordination number, one can construct a controlled approximation to treat both kinetic and interaction energy scales on the same footing.\cite{Metzner:prl1989, Georges:rmp1996} The basic idea in DMFT is to map the lattice problem onto a single (Anderson-type) impurity model embedded in a bath subjected to a self-consistency condition.  Although the general structure of the self-consistency loop is simple, solving the impurity problem is challenging in practice. In fact, in recent years much effort has been made to develop solvers for the impurity problem, leading to Exact Diagonalization (ED),\cite{Caffarel:prl1994} hybridization expansion,\cite{Werner:prl2006,Werner:prb2006} continuous time quantum Monte Carlo (QMC) methods,\cite{Gull:rmp2011} and also approximate methods such as the noncrossing approximation (NCA),\cite{Ruegg:prb13} renormalization group, slave-particle and perturbative methods.\cite{Georges:rmp1996} Each of these solvers has its own benefits and short comings.  While QMC suffers at zero temperature, ED is applicable at zero temperature but has finite-size issues.  Hence, both methods are computationally expensive.  Thus, resorting to approximate methods which are physically reasonable and computationally feasible is indispensable.  Among approximate methods, the iterative perturbation theory (IPT) is known to be reliable and simple to implement.\cite{Georges:prb1992, Rozenberg:prb1994, Rozenberg:prl1995,Matsuo:prb11}

In order to have a self-contained discussion, in the following we briefly review the formulation of DMFT for a single band, and its generalization for multi-band models.  Since the electrons in the bath are assumed to be noninteracting, the bath Green function is given by the hybridization function $\Delta(i\omega_n)$ as follows
\bea 
\label{g0} \mathcal{G}(i\omega_n)=\frac{1}{i\omega_n+\mu_0-\Delta(i\omega_n)},
\eea 
where $\omega_n$ is the Mutsubara frequency for fermions, and $\mu_0$ is a fictitious chemical potential of the bath which is determined by imposing sum rules or by other methods. On-site correlations exist on the impurity site leading to a nonzero local self-energy $\Sigma(i\omega_n)$. Hence, the Green function of the local impurity $f$ with onsite energy $\varepsilon_f$ is
\begin{equation}
 G^{-1}_f(i\omega_n)=\mathcal{G}^{-1}(i\omega_n)-\mu_0-\varepsilon_f-\Sigma(i\omega_n).
\end{equation}
  Moreover, the lattice Green function is defined as 
 \begin{equation} 
 G(i\omega_n)=\sum_{\bf{k}}\frac{1}{i\omega_n-\varepsilon_{\bf{k}}+\mu-\Sigma(i\omega_n)},
 \end{equation}
  where $\varepsilon_{\bf{k}}$ and $\mu$ are the dispersion of noninteracting electrons on the lattice and the chemical potential, respectively. The self-consistency condition connects the impurity and the lattice Green function by imposing $G(i\omega_n)=G_f(i\omega_n)$ and $\mu=-\varepsilon_f$. Therefore, we have a closed set of equations which must be iterated until numerical convergence is reached. At half-filling IPT is nothing but second order perturbation in the Hubbard $U$ as $\Sigma^{(2)}(i\omega_n)=-U^2\int_0^{\beta} e^{i\omega_n\tau}\mathcal{G}^2(\tau) \mathcal{G}(-\tau)$. Hence, the self-consistency loop can be easily performed. Note that we do not need to impose sum rules to fix the chemical potentials because at half-filling $\mu_0=0$ and $\mu=U/2$. One advantage of IPT is that it can be simply formulated in the real frequency domain by use of the spectral representation of Green functions,\cite{H.BARMAN:ijpb2011} namely $\mathcal{G}(i\omega_n)=-\frac{1}{\pi}\int_{-\infty}^{\infty}d\omega'\frac{\Im \mathcal{G}(\omega')}{i\omega_n-\omega'}$, so one is not required to perform an analytical calculation, which is a cumbersome and time consuming task. Hence, throughout we used a real frequency formulation of IPT. 

To go beyond half-filling and in order to address the doped systems, a modified version of IPT has been introduced and shown to be quantitatively accurate.\cite{Kajueter:prl1996} The scheme is based on an interpolative approach between various correct limits: the high frequency, atomic limit and the weakly interacting limit given by $\Sigma^{(2)}(\omega)$. The interpolated expression for the self-energy is\cite{Kajueter:prl1996} 
\bea 
\label{self_1orbital} 
\Sigma(\omega)=Un+\frac{A\Sigma^{(2)}(\omega)}{1-B\Sigma^{(2)}(\omega)}, 
\eea 
where $n$ is the filling, $A=\frac{n(1-n)}{n_0(1-n_0)}$, and $B=\frac{(1-n)U+\mu_0-\mu}{n_0(1-n_0)U^2}$  are functions of the chemical potentials via $n_0=-\frac{1}{\pi}\int_{-\infty}^{\infty}f(\omega)\Im{\mathcal{G}(\omega)}d\omega$ and $n=-\frac{1}{\pi}\int_{-\infty}^{\infty}f(\omega)\Im{G(\omega)}d\omega$.  Here $f(\omega)$ is the Fermi function. The quantities $\mu_0$ and $\mu$ are fixed by the filling and the Friedel sum rule\cite{Langreth:pr1966} or the Luttinger theorem.\cite{Luttinger:pr1960}  It is argued that by using the relation between double occupancy and the self-energy, $\langle n_{\uparrow}n_{\downarrow}\rangle=-\int_{-\infty}^{\infty}f(\omega)\Im{[\Sigma(\omega)G(\omega)]}d\omega/\pi U$, one can significantly improve the results in the strong coupling limit at finite temperature.\cite{Arsenault:prb2012} In this case, the results are in good agreement with essentially exact continuous time QMC.  Thus, our modified IPT theory solver is expected to be reliable over the full range of temperatures, frequencies, fillings, and Hubbard $U$ we consider.  For the single band model we use the self-energy relation in Eq.\eqref{self_1orbital} to calculate the interacting Green function $G(\omega)$ in self-consistency loop. 

The above interpolative scheme for the self-energy can be extended to the multi-band Hubbard model.\cite{Kotliar:prb1996} Doing so, the multi-orbital Hubbard model in Eq.\eqref{H} is mapped to a corresponding multi-orbital impurity model. On a single multi-orbital impurity in the atomic limit, where the coupling to the bath is zero, multiple charge states arise due to the inter-orbital interactions $V$ and $V'$. All charge states contribute to the structure of the atomic Green function, each with a weight given by many body correlations $\langle n_{m\sigma}n_{m'\sigma'}...\rangle$.\cite{Yeyati:prl1999} Without loss of generality and in order to keep the formalism as simple as possible, in the following we consider a two-band model. Having two orbitals, the Green function will have eight poles, and thus the interpolated self-energy in Eq.\eqref{self_1orbital} is generalized to a continuous fraction\cite{Kajueter:thesis} 
\bea
 \label{self_2orbital}
 \Sigma(\omega)&=&(U+V+V')n+\nonumber \\ &&\frac{C_1\Sigma^{(2)}(\omega)}{1+B_1\Sigma^{(2)}(\omega)+\frac{C_2(\Sigma^{(2)}(\omega))^2}{1+B_2\Sigma^{(2)}(\omega)+\frac{C_3(\Sigma^{(2)}(\omega))^2}{\frac{\ddots}{1+B_7\Sigma^{(2)}(\omega)}}}}, \nonumber \\ 
 \eea 
 where the coefficients $B_i$ and $C_i$ ($i=1..7$) depend on many body correlations and can be calculated from spectral moments or by use of an approximate method such as the coherent potential approximation.\cite{Kajueter:thesis} A more simplified version of the self-energy can be adopted assuming $U$ is sufficiently large that fluctuations in the mean charge of the impurity change by no more than one electron.\cite{Yeyati:prl1999} However, for maximal quantitative accuracy, in this work we consider the full expression for the self-energy given in Eq.\eqref{self_2orbital}.  

Having introduced the self-energy for the single and two orbital Hubbard models, we can close the self-consistency loop by following the steps: (i) start with a guess for the hybridization function $\Delta(\omega)$, (ii) calculate $\Sigma^{(2)}(\omega)$, (iii) calculate values of $\mu_0$ and $\mu$ for a fixed filling by use of sum rules and (iv) update the hybridization function as $\Delta(\omega)=\omega+\mu-G^{-1}(\omega)-\Sigma(\omega)$.            

Once the interacting Green function is calculated, we have all the necessary ingredients to evaluate the transport coefficients that govern the electrical and thermal responses of the model. They are given in terms of current-current correlation functions.\cite{mahan:book} The explicit expressions within the Kubo formalism are\cite{Tomczak:prb10}
\bea 
\label{Seebeck} 
&&S=-\frac{k_B}{|e|}\frac{A_1}{A_0}, \\ \label{conductivity} &&\sigma=\frac{2\pi e^2}{\hbar}A_0, \\ \label{thermal_con} &&\kappa=\kappa_l+\frac{2\pi k_B^2}{\hbar}T(A_2-\frac{A_1^2}{A_0}), 
\eea
where $A_0$ and $A_1$ are current-current and current-heat correlations, respectively. The correlations are given by\cite{Tomczak:prb10}
\bea 
\label{An} 
A_n=\int_{-\infty}^{\infty} d\omega \beta^n\omega^n\left(-\frac{\partial f}{\partial\omega}\right)\Xi(\omega).
\eea 
Since the self-energy is momentum independent, the transport kernel $\Xi(\omega)$ is given by 
\bea 
\Xi(\omega)=\int_{-\infty}^{\infty}d\varepsilon \rho(\omega,\varepsilon)^2 \Phi(\varepsilon),
\eea 
where the interaction effects are included in the spectral function as $\rho(\omega,\varepsilon)=-\frac{1}{\pi}\Im{\frac{1}{\omega-\varepsilon+\mu-\Sigma(\omega)}}$, and $\Phi(\varepsilon)$ is the transport function fully determined by the noninteracting electron dispersion as $\Phi(\varepsilon)=\frac{1}{V}\sum_{\bf{k}}\left(\frac{\partial\varepsilon_{\bf{k}}}{\partial k_x}\right)^2\delta(\varepsilon_{\bf{k}}-\varepsilon)$. Thus, these identifications make it clear how the interactions and information about the noninteracting band structure enter the transport coefficients. We calculated these functions numerically for each lattice we considered. Note that because of the non-interacting nature of $\Phi(\varepsilon)$, it should  only be calculated once. For simple cases, such as the hypercubic and Bethe lattices it can also be expressed analytically in terms of integrals of energy weighted by the non-interacting density of states.\cite{Arsenault:arxiv2013} 

\section{single band models\label{one_orbital}}
\begin{figure}[t]
\includegraphics[width=8cm]{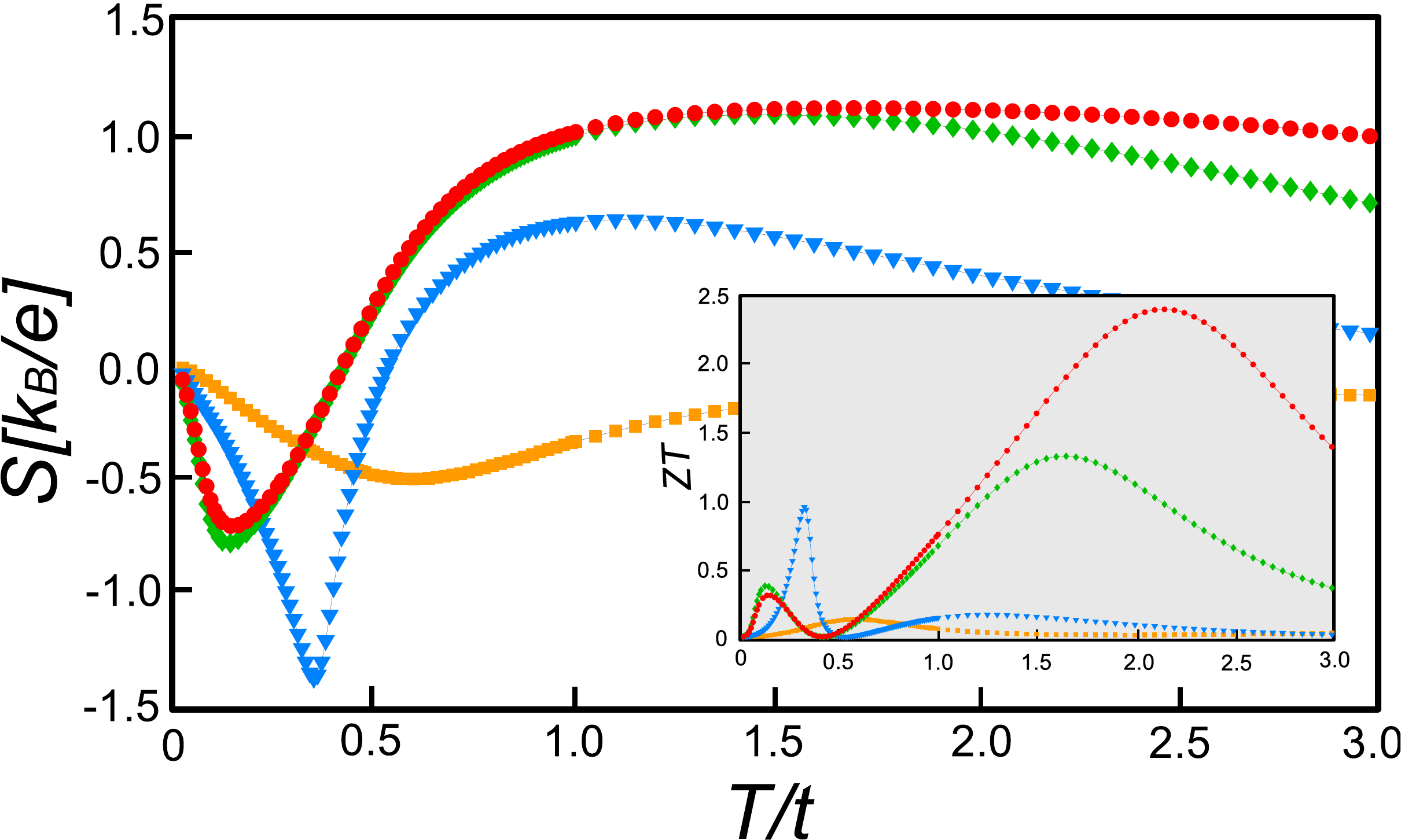}
\caption{(Color online) Seebeck coefficient as a function of temperature for the single-band square lattice at different values of $U=$ 4t (brown squares), 8t (blue down triangles), 14t (green lozenges) and 20t (red solid circles). The units are in fundamental constants $k_B$, $e$ and $\hbar$. Inset presents the figure-of-merit ${ZT}$. }\label{SO_Square}
\end{figure}

In this section we consider single band models on different lattices in two and three dimensions with only nearest neighbor hopping, denoted by $t$, and we measure all energy scales with respect to it. We start with the Hubbard model on the square lattice whose noninteracting dispersion is $\varepsilon_{\bf{k}}=-2t(\cos(k_x)+\cos(k_y))$ with lattice constant $a=1$. The filling fraction is set to $n=0.85$ corresponding to a (hole) doping $\delta=0.15$ away from half-filling (where $n=1$). Thus, the system is in strong correlated regime. For dilute electron density corresponding to lower fillings the transport properties are mainly driven by noninteracting electrons.\cite{Arsenault:prb13} Fig.\ref{SO_Square} displays the variation of the Seebeck coefficient and the figure-of-merit with respect to temperature at different values of the Hubbard interaction $U$. We do not consider the phonon contribution to the thermal conductivity; we only consider the contribution of charge carriers to thermal conductivity, so hereafter we take $\kappa_l=0$. 

In Fig.\ref{SO_Square} it is clearly seen that the correlations enhance the Seebeck coefficient over a wide range of temperatures. In particular, the Seebeck coefficient is maximized when the value of $U$ is of order of the band width. For a square lattice the band with is $W=8t$ and the maximum occurs at $U/W \sim 1$. This is an interesting result--for cuprates $U\approx 4.0 ~\mathrm{eV}$ and $t\approx 0.5 ~\mathrm{eV}$, which then puts them in a category of materials with large Seebeck coefficient.\cite{Chakraborty:prb10} The variation of the Seebeck coefficient at $U=14t$ and $U=20t$ is particularly interesting as the system is a Mott insulator at half filling. It shows that the anomalous behavior of the transport functions can be described reasonably well by quasiparticles with a quadratic temperature dependence surviving well above the Fermi temperature scale.\cite{Xu:prl2013} We should also emphasize that the values of the Seebeck coefficient for $U=14t$ are in agreement with the corresponding values calculated by use of CTQMC in Ref.[\onlinecite{Xu:prl2013}], which shows that the modified IPT yields a fairly accurate description of this correlated system. The thermal conductivity decreases with increasing correlations over entire range of temperatures as shown in Fig.\ref{thermal_con}. This reduction give rises to a large figure-of-merit at both low and high temperatures. It becomes zero at the temperature where the Seebeck coefficient changes sign. The sign change of the Seebeck coefficient is an interaction effect not present in the noninteracting ($U=0$) electron model.

\begin{figure}[t]
\includegraphics[width=8cm]{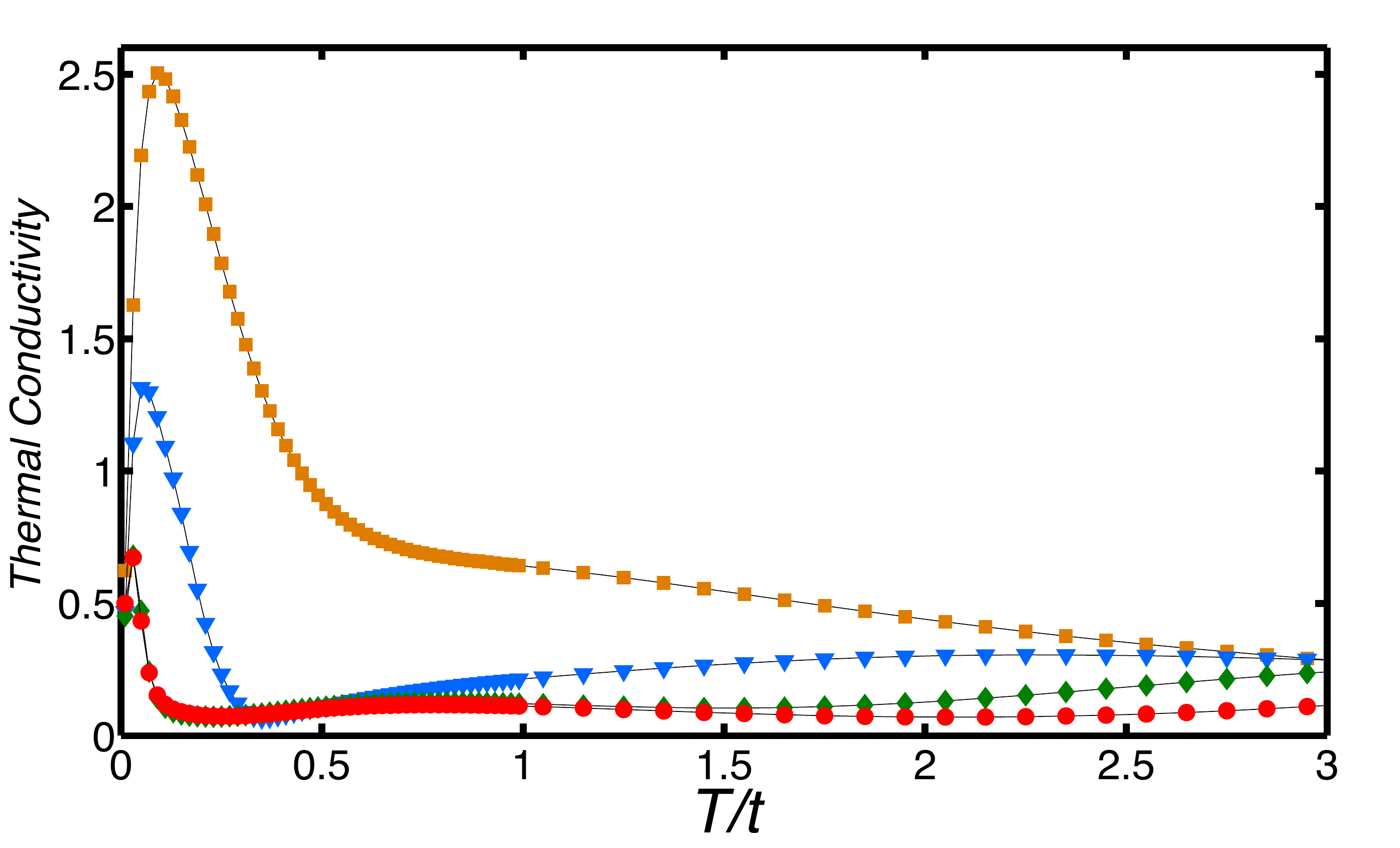}
\caption{(Color online) Electron thermal conductivity $\kappa_e$ in unite of $k_B^2/\hbar$ as a function of temperature for the single-band square lattice at different values of $U=$ 4t (brown squares), 8t (blue down triangles), 14t (green lozenges) and 20t (red solid circles). }\label{thermal_con}
\end{figure} 

\begin{figure}[t]
\includegraphics[width=8cm]{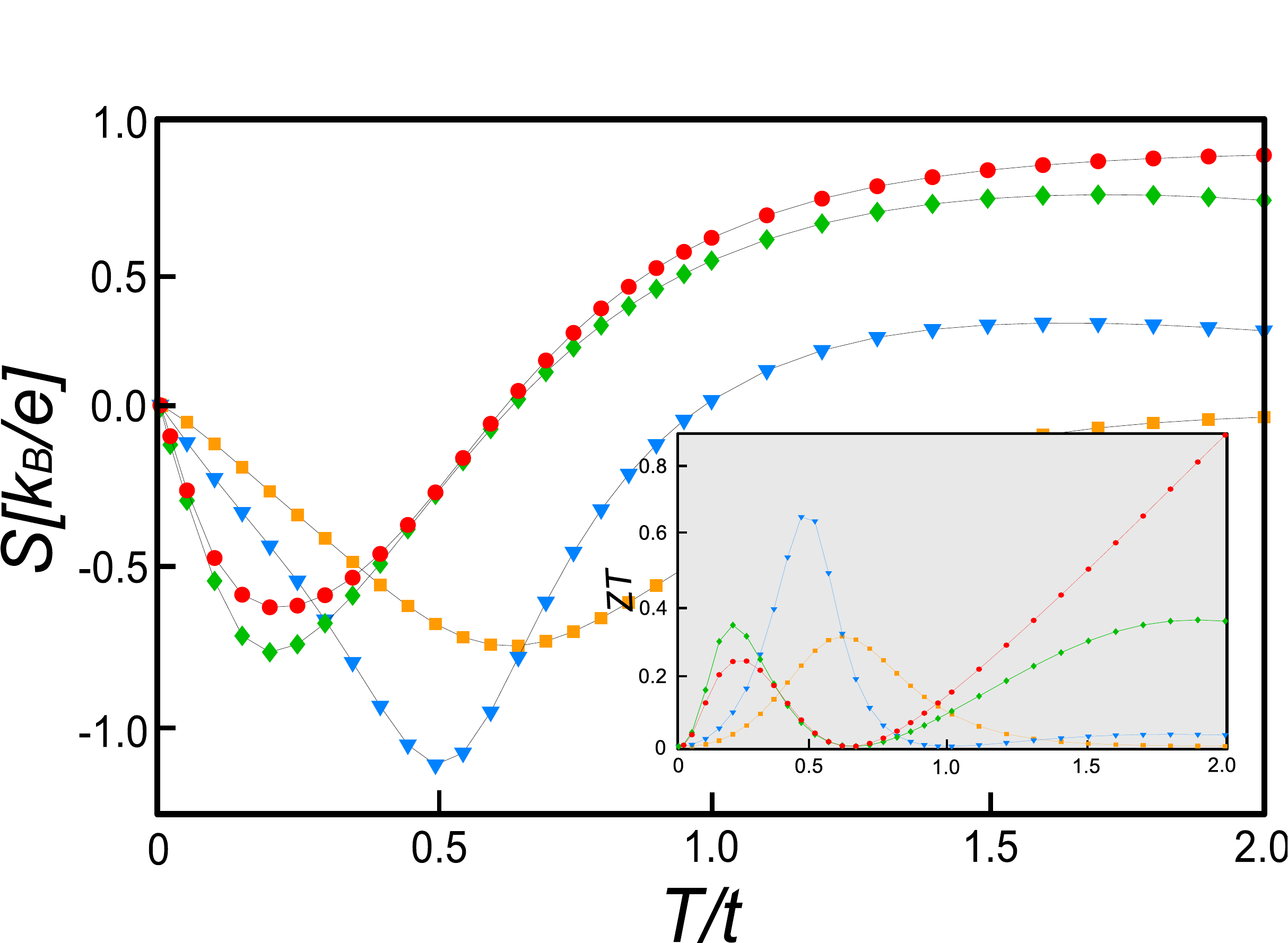}
\caption{(Color online) Seebeck coefficient as a function of temperature for the single-band simple cubic lattice at different values of $U=$ 6t (brown squares), 8t (blue down triangles), 12t (green lozenges) and 24t (red solid circles). The units are in fundamental constants $k_B$, $e$ and $\hbar$. Inset presents the figure of merit ${ZT}$. }\label{SO_Cubic}
\end{figure}

We now turn to three dimensional systems. We first consider the simple cubic lattice. The noninteracting dispersion is $\varepsilon_{\bf{k}}=-2t(\cos(k_x)+\cos(k_y)+\cos(k_z))$, and we set the (hole) doping level to $\delta=0.2$ ($n=0.8$) away from half-filling. We studied the Hubbard model in the full range of interactions from weak to strong. The variation of the Seebeck coefficient and ${ZT}$ are shown in Fig.\ref{SO_Cubic}. While at low temperatures the Seebeck coefficient is strongly temperature dependent, it  saturates at higher temperatures.  Similar to the square lattice, electronic correlations tend to increase the Seebeck coefficient at most temperatures, and at low temperatures the maximum is reached for values of $U$ close to the band width, which then leads to a high figure-of-merit (see inset in Fig.\ref{SO_Cubic}). As far as high temperature applications are concerned, correlations larger than the band width yield a higher figure-of-merit than the intermediate or weakly correlated states.  Note that at low temperatures the value of $U$ where the maximum of the Seebeck coefficient occurs generally depends on the doping level, but we found for a range of doping close to half filling the maximum is reached for $U\sim W$. 

\begin{figure}[t]
\includegraphics[width=8cm]{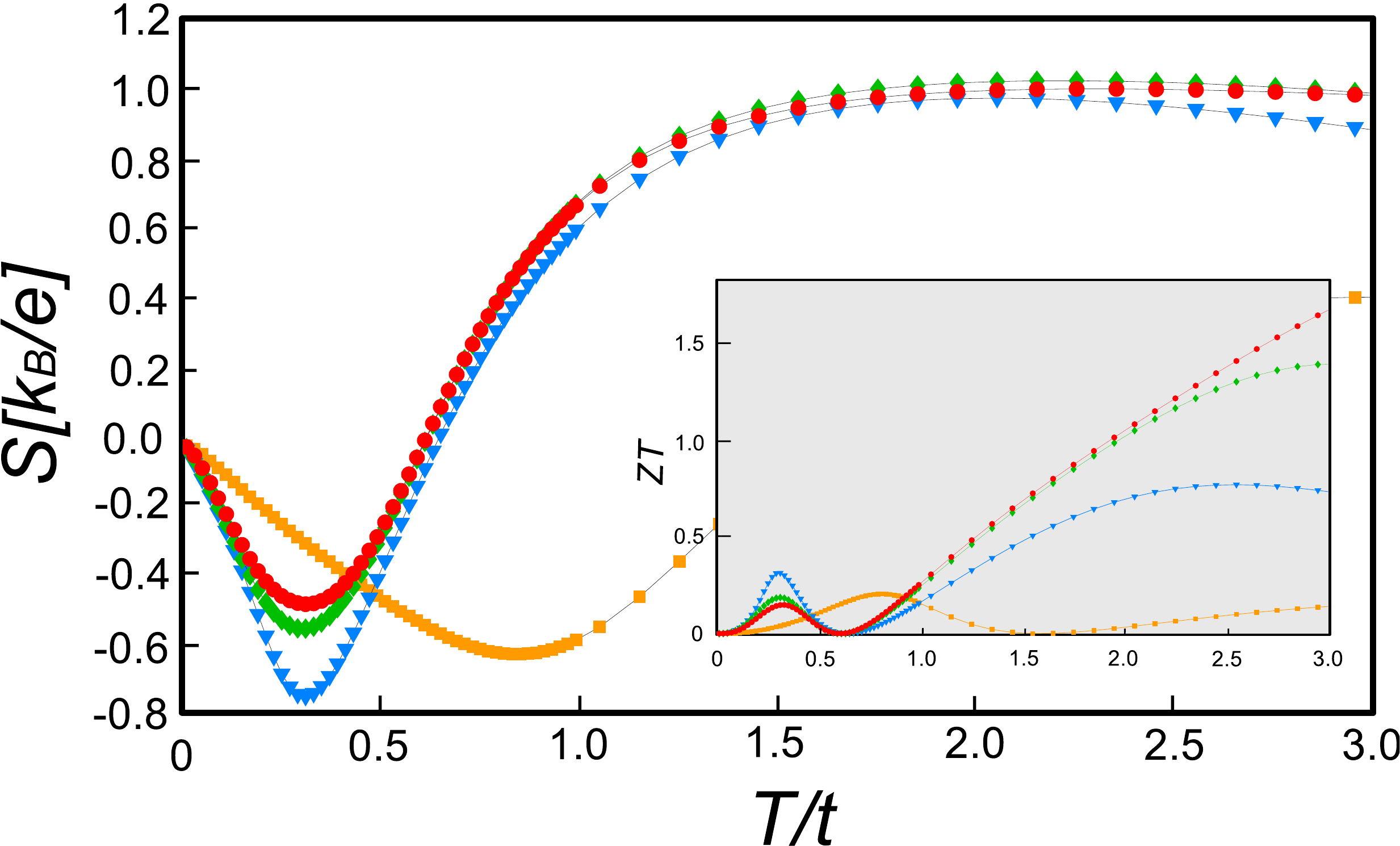}
\caption{(Color online) Seebeck coefficient as a function of temperature for the single-band face center cubic lattice at different values of $U=$ 8t (brown squares), 16t (blue down triangles), 24t (green lozenges) and 32t (red solid circles). The units are in fundamental constants $k_B$, $e$ and $\hbar$. Inset presents the figure of merit ${ZT}$. }\label{SO_FCC}
\end{figure}

We also studied the single band model on the face centered cubic (FCC) lattice with non-interacting dispersion $\varepsilon_{\bf{k}}=-4t(\cos(k_x/2)\cos(k_y/2)+\cos(k_y/2)\cos(k_z/2)+\cos(k_x/2)\cos(k_z/2))$.  We chose an electron density corresponding to filling $n=0.8$. The results are shown in Fig.\ref{SO_FCC}. We find a broad maximum in the magnitude of the Seebeck coefficient which becomes more pronounced as $U$ increases. It has been argued that in the presence of large enough interactions, corresponding to an insulator phase at half-filling, large thermoelectric effects appear at low temperatures.\cite{Arsenault:prb13} We, however, show that, similar to the thermoelectric properties of square and cubic lattices discussed above, the largest Seebeck coefficient occurs for a less correlated state where the interaction is of order of the band width of the noninteracting dispersion. It is clearly seen that the largest Seebeck coefficient is achieved when the $U=16t$, which corresponds to $U/W\sim 1$ (note W=16t for the FCC lattice). 
             
\section{two band models\label{two_orbital}}
In this section we consider the model Hamiltonian corresponding to Eq.\eqref{H} on the lattices studied in the single-band approximation in the preceding section. We consider two degenerate orbitals on each site of the lattice and for the sake of simplicity we assume that the hopping matrix is diagonal in the orbital basis, namely $t_{mm'}=t\delta_{mm'}$ implying that there is no direct overlap between orbitals. Of course, the electrons on different orbitals interact via the couplings $V$ and $V'$, which alters the spectral density compared to the one band model because of multiple excited states. The latter excitations introduce new poles in the Green function and make the self energy quite complicated, as in Eq.\eqref{self_2orbital}. In the following, we study the effect of multiple excitations on the thermoelectric properties of electronic systems and compare them with the one-band models studied in the previous section.    

\begin{figure}[t]
\includegraphics[width=8cm]{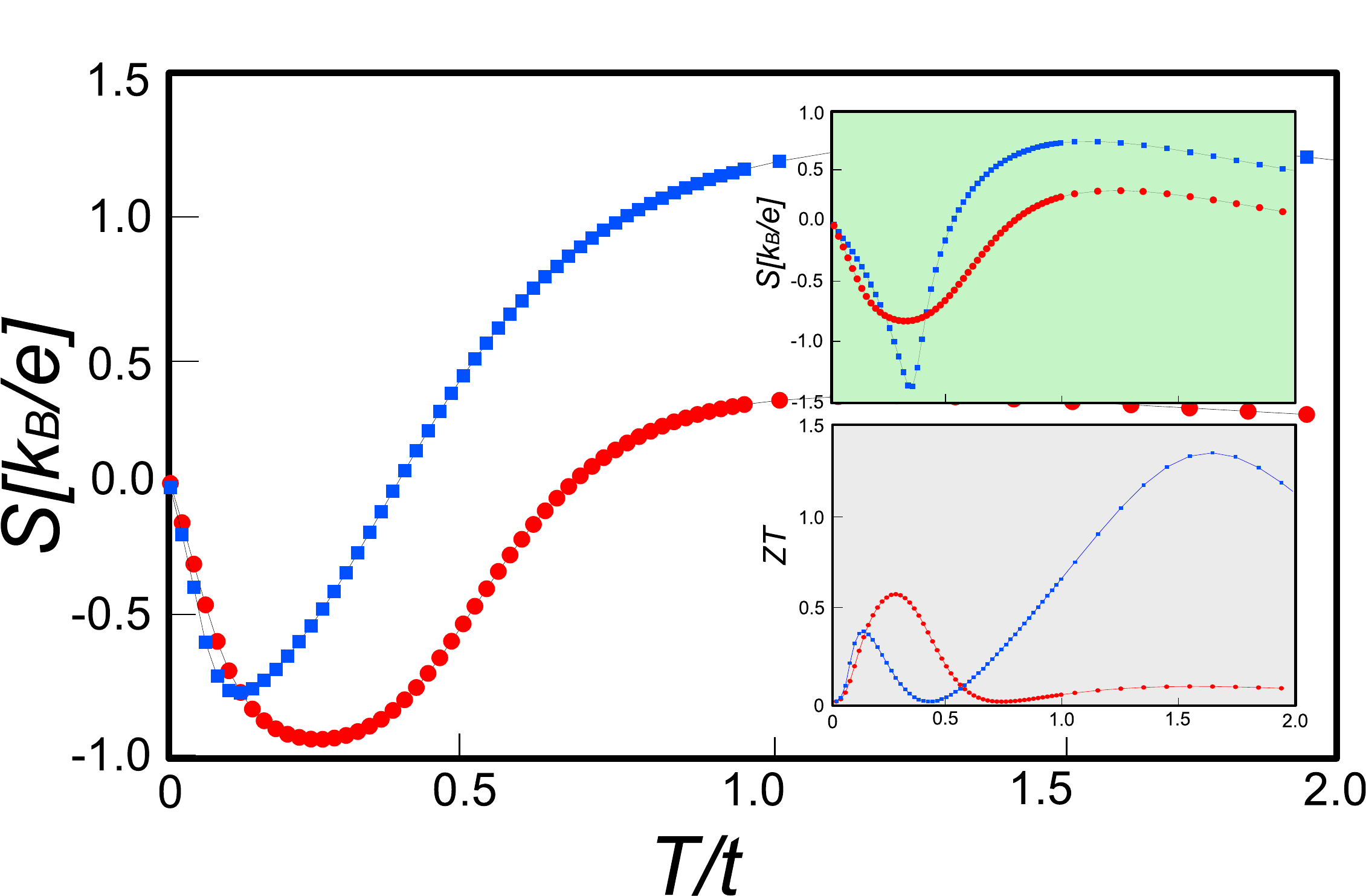}
\caption{(Color online) Seebeck coefficient as a function of temperature for one-band and two-band square lattice models.  The main plot indicates the results for $U=14t$: Red solid circles and blue squares stand for two and single band model, respectively. Top inset: Seebeck coefficient versus temperature for $U=8t$ and color coding (symbols) is the same as main plot. Bottom inset: $ZT$ versus temperature for $U=14t$ and color coding (symbols) is the same as main plot. }\label{2iO_Square}
\end{figure}

Fig.\ref{2iO_Square} shows the results for the square lattice. We consider weakly and strongly correlated cases corresponding to $U=8t$ and $U=14t$, respectively. The main plot indicates the comparison with the single-orbital model for strong correlation $U=14t$: red (circles) and blue (squares) symbols indicate the variation of the Seebeck coefficient for the model with two and one orbital, respectively.  The top inset shows the same comparison, but for $U=8t$. Much like the one-orbital model, the correlations increase the Seebeck coefficient at low temperatures. Note that at this specific doping and at low temperatures, the values of the Seebeck coefficient for the correlated system with $U=14t$ are larger than the corresponding values for $U=8t$ in the two band model. This is in contrast with the single-orbital case in the preceding section. Moreover, while at very low temperatures, say $T<0.1t$, the value of Seebeck is almost the same for both models, it is larger for the two orbital model at higher temperatures $0.1t<T<0.5t$. This is also clear from the figure-of-merit data shown in inset on bottom of Fig.\ref{2iO_Square}. For intermediate correlations, as shown in the inset on top, the Seebeck coefficient of the two-orbital model is smaller. Hence, it seems that the Seebeck coefficient only exceeds the one-orbital model (at lower temperatures) when the interactions are strong enough. However, at high temperatures $T>0.5t$ the value of Seebeck coefficient for the one-orbital is found to be generally larger than the two-orbital model.  Correspondingly, as seen in the bottom inset of Fig.\ref{2iO_Square}, the value of $ZT$ of the one-orbital model is much larger than that of two-orbital model at high temperatures. 

\begin{figure}[t]
\includegraphics[width=8cm]{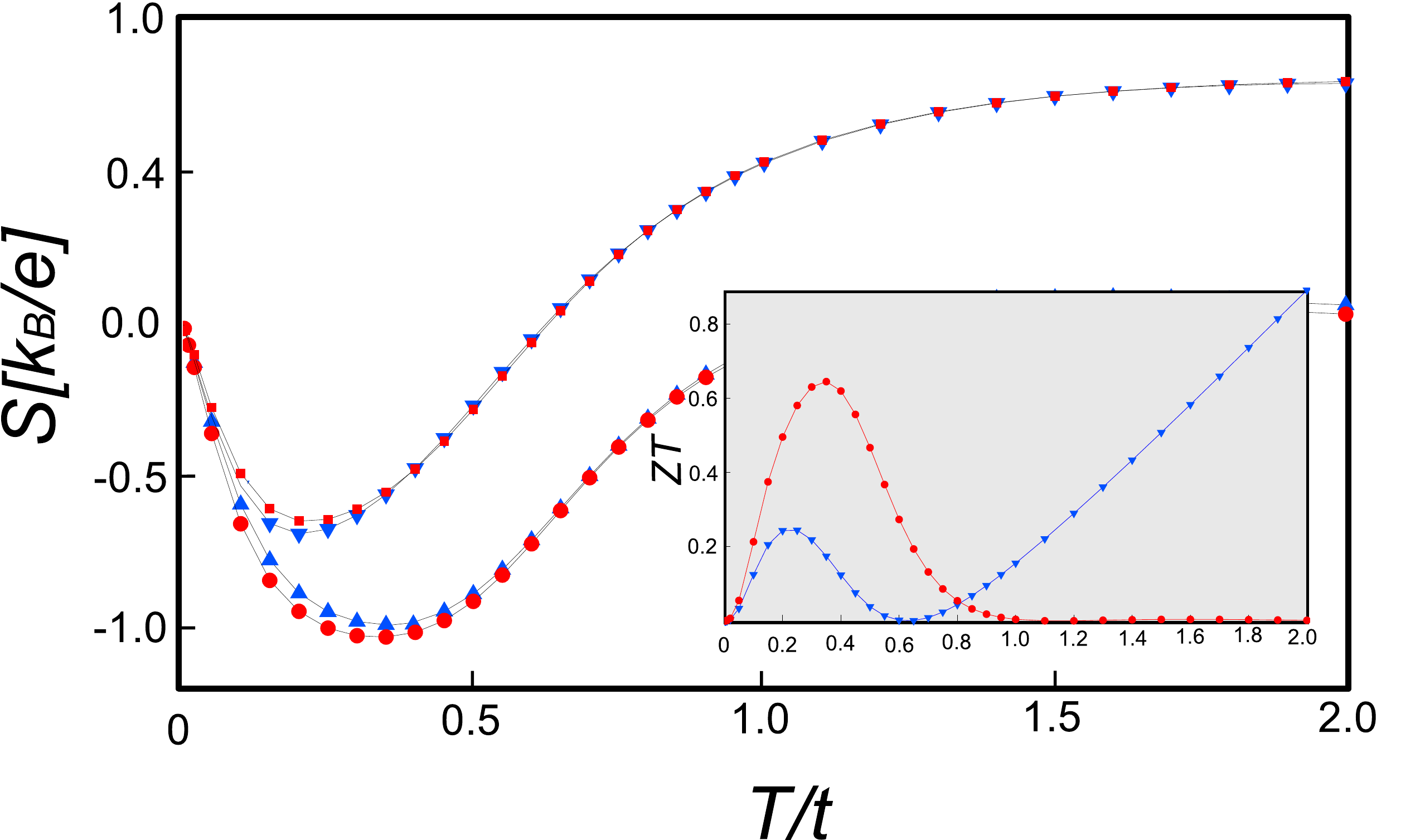}
\caption{(Color online) Seebeck coefficient as a function of temperature for cubic lattice. Red squares and blue down triangles encode data at different values of $U=$ 18t (down triangles) and 24t (squares) of a single-band model. Plots in bottom encode the data for two band model at different values of $U=$ 18t (blue up triangles) and 24t (red solid circles). In the inset we indicate the figure-of-merit ${ZT}$ only for $U=24t$: red solid circles and blue triangles for two and one band models, respectively. }\label{2iO_Cubic}
\end{figure}

We next compare the Seebeck coefficient of the two-orbital model with the single-orbital model on the cubic lattice.  Our results are shown in Fig.\ref{2iO_Cubic}. Blue down triangles (blue up triangles) and red squares (red solid circles) symbols show the data for single-orbital (two-orbital) model at $U=18t$ and $U=24t$, respectively. Unlike the evolution of the Seebeck coefficient in the single-orbital model where the maximum is reached at values of $U$ comparable to band width, in the two-orbital model the Seebeck coefficient is larger for weak correlations. This phenomenon indicates that the multi-excitations due to the multi-orbital nature of the system could play an important role in enhancing thermoelectric properties at low temperatures. In a wide range of temperatures $T \lesssim 0.7t$ the two-orbital model yields a (slightly) larger Seebeck coefficient than the single-band model. However, at higher temperatures both the Seebeck coefficient and ${ZT}$ (see inset in Fig.\ref{2iO_Cubic}) of the single-orbital model become larger than the corresponding values of two-orbital model.  Nevertheless, over a range of temperatures, $0 \lesssim T \lesssim 0.8t$, the figure-of-merit is significantly enhanced for the two-band model.  

\begin{figure}[t]
\includegraphics[width=8cm]{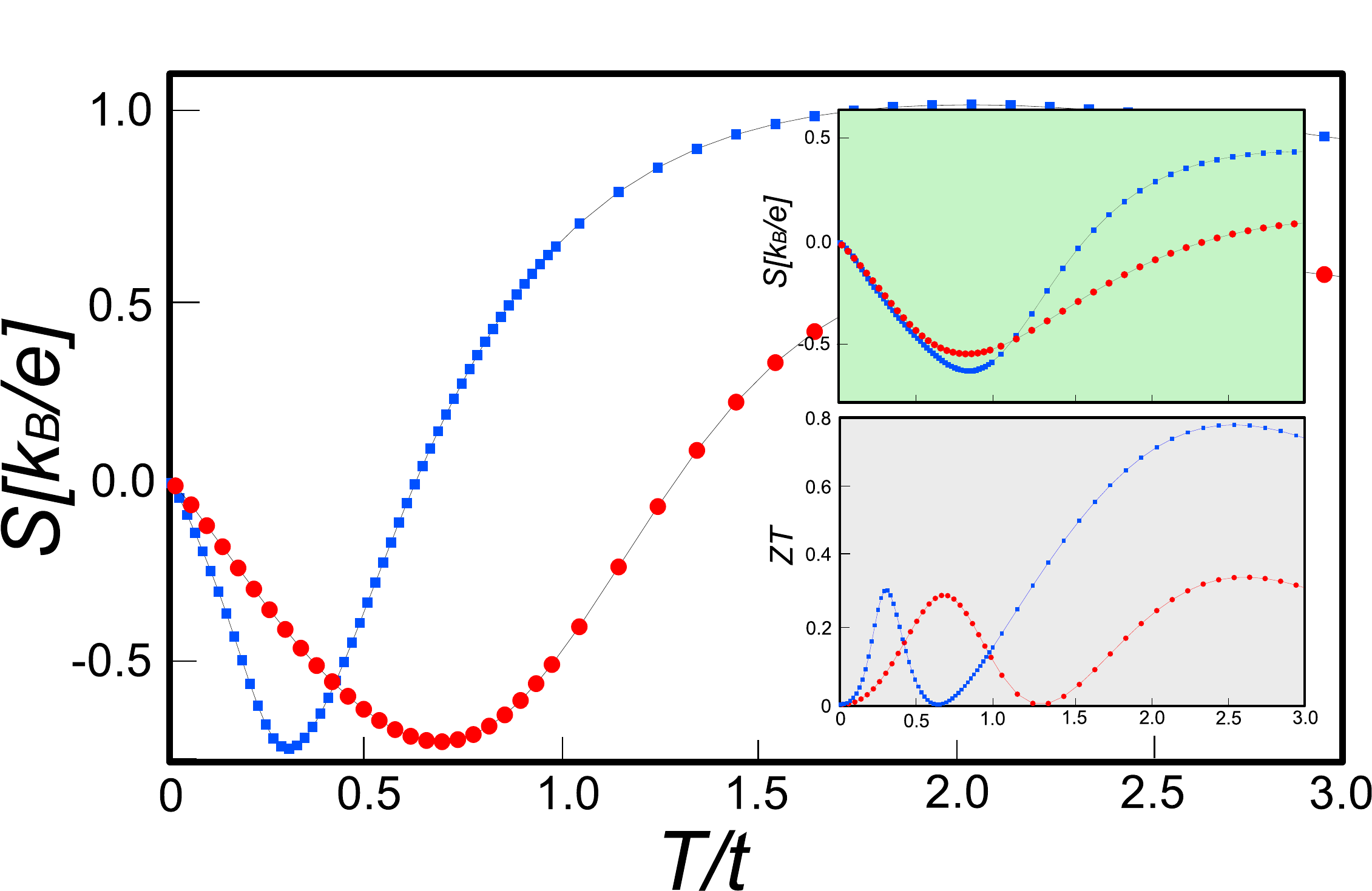}
\caption{(Color online) Seebeck coefficient as a function of temperature for the face-centered cubic lattice.  The main plot indicates the results for $U=16t$: Red solid circles and blue squares stand for the two and single-band models, respectively. Top inset: Seebeck coefficient versus temperature for $U=8t$ and color coding (symbols) is the same as main plot. Bottom inset: $ZT$ versus temperature for $U=16t$ and color coding (symbols) is the same as the main plot.}\label{2iO_FCC}
\end{figure}

Fig.\ref{2iO_FCC} compiles the same set of data on the FCC lattice. Blue squares and red solid circles symbols in main and also in insets stand for the one-orbital and the two-orbital model, respectively. The main plot presents the data for $U=16t$ and the top inset includes the data for $U=8t$.  At low temperatures, $T\lesssim 0.4t$, the single-band model yields a larger Seebeck coefficient than the two band model. But in a temperature window $0.4t\lesssim T \lesssim 1.0t$ the values of the coefficient for the two-orbital model exceeds the single-band model.  As the temperature rises both the Seebeck coefficient and ${ZT}$ (see bottom inset in Fig.\ref{2iO_FCC}) for the single-orbital model shows higher values than the two-orbital model.   

Taking all the lattice models together, if we estimate $t\sim eV$, then we expect significantly enhanced figure-of-merits for multi-orbital models in the temperature ranges of a few hundred Kelvin, which may be practical for applications.  When spin-orbit ad crystal field effects are taken into account, it appears that ``sweet spots" of enhanced figure-of-merit appear in the parameter space of multi-orbital models.\cite{Hong:prb13}

\section{Conclusions and discussion}\label{conclusions}
In this work we presented results for thermal transport properties for a variety of strongly correlated electron models on different lattices. In particular, we found the Seebeck coefficient (which measures the drop of voltage across a system due to a temperature gradient) in general has a non-monotonic temperature dependence, and even changes sign as a function of temperature.  In order to calculate the figure-of-merit ${ZT}$, we also calculated other transport coefficients such as the electrical and thermal conductivities. We used a multi-orbital Hubbard model as a prototype to address the effect of correlations and orbital degeneracy on the transport coefficients. We used DMFT with a modified IPT as the impurity solver to calculate the single-particle interacting Green function. The spectral density is then used to calculate the appropriate transport coefficients for both the single and two-orbital models. 

The results presented in Secs. \ref{one_orbital} and \ref{two_orbital}, though for different lattices, have some common features. For all cases the Seebeck coefficient shows a non-monotonic behavior with temperature. The same non-monotonic behavior is observed for Bethe lattice.\cite{Sekino:arxiv1308.0416} At very low temperatures the Seebeck coefficient is almost linear with respect to temperature. This is expected as low the temperature expansion of Eq.\eqref{An} yields an expression for the Seebeck coefficient linear in $T$. In this limit the Seebeck coefficient can be interpreted as the logarithmic derivative of the transport function with respect to the effective chemical potential, \cite{Palsson:prl1998} 
\bea 
\label{seebeck_expansion} 
S\simeq -\frac{k_B}{e}\left(\frac{k_BT}{Z}\frac{d\ln\Phi(\varepsilon)}{d\varepsilon}\right)_{\varepsilon=\tilde{\mu}},
\eea
 where $\tilde{\mu}=\mu-\Re\Sigma(0)$ is effective chemical potential and $Z$ is quasiparticle residue. Moreover, as all plots suggest, the slope of Seebeck coefficient at zero temperature increases with interaction. This is already evident from the dependence of the Seebeck coefficient on the quasiparticle residue $Z$: the smaller $Z$ due to correlations, the larger the Seebeck coefficient and the figure-of-merit. 

With increasing temperature the Seebeck coefficient becomes more negative and passes through a minimum. The appearance of such minimum is seen in all correlated cases we studied. The temperature at which this minimum appears, $T_{min}$, approximately corresponds to the temperature where a peak appears in the specific heat.\cite{Merino:prb2000} Indeed, the appearance of a minimum in the thermopower is a signature of the thermal destruction of the coherent Fermi liquid state that exists at low temperatures or perhaps a crossover to state with resilient quasiparticle.\cite{Deng:prl13} This latter phase is still characterized by well defined quasiparticle excitations but with pronounced particle-hole asymmetry between electron-like and hole-like excitation lifetimes which make the transport properties of the system anomalous.\cite{Zlatic:prl12} The values of $T_{min}$ also shift to lower temperatures by enhancing the interactions. Due to correspondence between the minimum in the Seebeck coefficient and the peak in the specific heat, $T_{min}$ signals the Kondo scale, {\em i.e} $T_{min}\sim T_{\bf{k}}$. The Kondo temperature $T_{\bf{k}}$ scales exponentially with interaction $U$ as $T_{\bf{k}}=\sqrt{\frac{2\Delta U}{\pi^2}}e^{-\frac{\pi U}{8\Delta}}$,\cite{Haldane:prl1978, Okiji:prl1983, Kawakami:1983} where $\Delta$ is the resonance width depending on the hybridization of the impurity to the bath $V_{\bf{k}}$ as $\Delta=\pi\sum_{\bf{k}}|V_{\bf{k}}|^2\delta(\varepsilon_{\bf{k}}-\mu)$. Thus, it is seen that for strong correlations $T_{\bf{k}}$ shifts to lower temperatures.  Additional arguments based on thermodynamic considerations appear to explain some aspects of qualitative trends, but not numerical values or details.\cite{Zlatic2013}

Besides the appearance of a minimum in the Seebeck coefficient and its shift to lower temperatures, the change of its sign at some intermediate temperature is also a common feature of all correlated states. In non-interacting systems, the sign of the Seebeck coefficient is determined by the type of carriers dominating the transport properties. However, in the presence of correlations the Seebeck coefficient can change sign.\cite{Pruschke:AdPhys1995} At low temperatures the transport properties are mainly dominated by electrons. By increasing the temperature, the spectral weight is transferred to the lower Hubbard band making the holes the dominant carriers. At high temperature the quasiparticle peak has almost disappeared and the incoherent carriers dominate the transport. The change in the sign of the Seebeck coefficient could also occur at a fixed temperature by changing the doping. In the latter case the particle-hole symmetry is restored upon doping in the presence of strong correlations.\cite{Chakraborty:prb10} At the particle-hole symmetric point the Seebeck coefficient vanishes due to equal contribution of both types of carriers. Thus, the Seebeck coefficient will have opposite sign at slightly above or below the doping at which the particle-hole symmetry is restored. In almost all plots it is seen that the Seebeck coefficient becomes saturated at values of order $k_B/e$ at high temperatures where the incoherent regimes sets in. In this limit the Seebeck coefficient essentially measures the entropy per particle and is independent of temperature.\cite{Chaikin:prb1976}  We have verified that the Kelvin formula\cite{Peterson:prb10} qualitatively captures the trends with temperature and interactions in the Hamiltonians we study, but it is numerically rather poor except at very low and very high temperatures.

We also considered the effect of explicitly including a Hunds coupling $J$ to Eq.(1) and found that it slightly increased the Seebeck coefficient at temperatures less the hopping $t$, but tended to decrease it at higher temperatures. The overall features, such as the sign change with temperature and dip for $T<t$, remained unchanged.

Hence, the anomalous behavior of the Seebeck coefficient and also the other thermoelectric coefficients can be traced back to different regimes of the metallic phase. Indeed, the transport properties of the correlated sates are modified due to passing through two crossovers: At very low temperatures, the Fermi liquid with well defined quasiparticles dominates the transport. Increasing the temperature drives the system to an intermediate metallic phase where the quasiparticles are still well defined but with an anomalous scattering rate evading Fermi liquid theory.\cite{Deng:prl13, Xu:prl2013} The quasiparticles of this latter metallic phase then become completely incoherent at high temperatures. 
\\\\
                                        
\section{acknowledgements}
M.K. would like to thank H. Barman and  L.-F. Arsenault for useful discussions. We gratefully acknowledge financial support from ARO Grant No. W911NF-09-1-0527,  NSF Grant No. DMR-0955778, and DARPA grant D13AP00052.


\begin{thebibliography}{59}%
\makeatletter
\providecommand \@ifxundefined [1]{%
 \@ifx{#1\undefined}
}%
\providecommand \@ifnum [1]{%
 \ifnum #1\expandafter \@firstoftwo
 \else \expandafter \@secondoftwo
 \fi
}%
\providecommand \@ifx [1]{%
 \ifx #1\expandafter \@firstoftwo
 \else \expandafter \@secondoftwo
 \fi
}%
\providecommand \natexlab [1]{#1}%
\providecommand \enquote  [1]{``#1''}%
\providecommand \bibnamefont  [1]{#1}%
\providecommand \bibfnamefont [1]{#1}%
\providecommand \citenamefont [1]{#1}%
\providecommand \href@noop [0]{\@secondoftwo}%
\providecommand \href [0]{\begingroup \@sanitize@url \@href}%
\providecommand \@href[1]{\@@startlink{#1}\@@href}%
\providecommand \@@href[1]{\endgroup#1\@@endlink}%
\providecommand \@sanitize@url [0]{\catcode `\\12\catcode `\$12\catcode
  `\&12\catcode `\#12\catcode `\^12\catcode `\_12\catcode `\%12\relax}%
\providecommand \@@startlink[1]{}%
\providecommand \@@endlink[0]{}%
\providecommand \url  [0]{\begingroup\@sanitize@url \@url }%
\providecommand \@url [1]{\endgroup\@href {#1}{\urlprefix }}%
\providecommand \urlprefix  [0]{URL }%
\providecommand \Eprint [0]{\href }%
\@ifxundefined \urlstyle {%
  \providecommand \doi  [0]{\begingroup \@sanitize@url \@doi}%
  \providecommand \@doi [1]{\endgroup \@@startlink {\doibase
  #1}doi:\discretionary {}{}{}#1\@@endlink }%
}{%
  \providecommand \doi  [0]{doi:\discretionary{}{}{}\begingroup
  \urlstyle{rm}\Url }%
}%
\providecommand \doibase [0]{http://dx.doi.org/}%
\providecommand \Doi [0]{\begingroup \@sanitize@url \@Doi }%
\providecommand \@Doi  [1]{\endgroup\@@startlink{\doibase#1}\@@Doi}%
\providecommand \@@Doi [1]{#1\@@endlink}%
\providecommand \selectlanguage [0]{\@gobble}%
\providecommand \bibinfo  [0]{\@secondoftwo}%
\providecommand \bibfield  [0]{\@secondoftwo}%
\providecommand \translation [1]{[#1]}%
\providecommand \BibitemOpen [0]{}%
\providecommand \bibitemStop [0]{}%
\providecommand \bibitemNoStop [0]{.\EOS\space}%
\providecommand \EOS [0]{\spacefactor3000\relax}%
\providecommand \BibitemShut  [1]{\csname bibitem#1\endcsname}%
\bibitem [{\citenamefont {Mahan}\ \emph {et~al.}(1997)\citenamefont {Mahan},
  \citenamefont {Sales},\ and\ \citenamefont {Sharp}}]{Mahan:phyicstoday97}%
  \BibitemOpen
  \bibfield  {author} {\bibinfo {author} {\bibfnamefont {G.}~\bibnamefont
  {Mahan}}, \bibinfo {author} {\bibfnamefont {B.}~\bibnamefont {Sales}}, \ and\
  \bibinfo {author} {\bibfnamefont {J.}~\bibnamefont {Sharp}},\ }\href@noop {}
  {\bibfield  {journal} {\bibinfo  {journal} {Physics Today},\ }\textbf
  {\bibinfo {volume} {50}},\ \bibinfo {pages} {42} (\bibinfo {year}
  {1997})}\BibitemShut {NoStop}%
\bibitem [{\citenamefont {Snyder}\ and\ \citenamefont
  {Toberer}(2008)}]{Snyder:nmat08}%
  \BibitemOpen
  \bibfield  {author} {\bibinfo {author} {\bibfnamefont {G.~J.}\ \bibnamefont
  {Snyder}}\ and\ \bibinfo {author} {\bibfnamefont {E.~S.}\ \bibnamefont
  {Toberer}},\ }\href {http://dx.doi.org/10.1038/nmat2090} {\bibfield
  {journal} {\bibinfo  {journal} {Nat Mater},\ }\textbf {\bibinfo {volume}
  {7}},\ \bibinfo {pages} {105} (\bibinfo {year} {2008})}\BibitemShut {NoStop}%
\bibitem [{\citenamefont {Poudel}\ \emph {et~al.}(2008)\citenamefont {Poudel},
  \citenamefont {Hao}, \citenamefont {Ma}, \citenamefont {Lan}, \citenamefont
  {Minnich}, \citenamefont {Yu}, \citenamefont {Yan}, \citenamefont {Wang},
  \citenamefont {Muto}, \citenamefont {Vashaee}, \citenamefont {Chen},
  \citenamefont {Liu}, \citenamefont {Dresselhaus}, \citenamefont {Chen},\ and\
  \citenamefont {Ren}}]{Poudel:Sceince2008}%
  \BibitemOpen
  \bibfield  {author} {\bibinfo {author} {\bibfnamefont {B.}~\bibnamefont
  {Poudel}}, \bibinfo {author} {\bibfnamefont {Q.}~\bibnamefont {Hao}},
  \bibinfo {author} {\bibfnamefont {Y.}~\bibnamefont {Ma}}, \bibinfo {author}
  {\bibfnamefont {Y.}~\bibnamefont {Lan}}, \bibinfo {author} {\bibfnamefont
  {A.}~\bibnamefont {Minnich}}, \bibinfo {author} {\bibfnamefont
  {B.}~\bibnamefont {Yu}}, \bibinfo {author} {\bibfnamefont {X.}~\bibnamefont
  {Yan}}, \bibinfo {author} {\bibfnamefont {D.}~\bibnamefont {Wang}}, \bibinfo
  {author} {\bibfnamefont {A.}~\bibnamefont {Muto}}, \bibinfo {author}
  {\bibfnamefont {D.}~\bibnamefont {Vashaee}}, \bibinfo {author} {\bibfnamefont
  {X.}~\bibnamefont {Chen}}, \bibinfo {author} {\bibfnamefont {J.}~\bibnamefont
  {Liu}}, \bibinfo {author} {\bibfnamefont {M.~S.}\ \bibnamefont
  {Dresselhaus}}, \bibinfo {author} {\bibfnamefont {G.}~\bibnamefont {Chen}}, \
  and\ \bibinfo {author} {\bibfnamefont {Z.}~\bibnamefont {Ren}},\ }\href
  {http://www.sciencemag.org/content/320/5876/634.abstract} {\bibfield
  {journal} {\bibinfo  {journal} {Science},\ }\textbf {\bibinfo {volume}
  {320}},\ \bibinfo {pages} {634} (\bibinfo {year} {2008})}\BibitemShut
  {NoStop}%
\bibitem [{\citenamefont {Cutler}\ \emph {et~al.}(1964)\citenamefont {Cutler},
  \citenamefont {Leavy},\ and\ \citenamefont {Fitzpatrick}}]{Cutler:prb1964}%
  \BibitemOpen
  \bibfield  {author} {\bibinfo {author} {\bibfnamefont {M.}~\bibnamefont
  {Cutler}}, \bibinfo {author} {\bibfnamefont {J.~F.}\ \bibnamefont {Leavy}}, \
  and\ \bibinfo {author} {\bibfnamefont {R.~L.}\ \bibnamefont {Fitzpatrick}},\
  }\href {http://link.aps.org/doi/10.1103/PhysRev.133.A1143} {\bibfield
  {journal} {\bibinfo  {journal} {Physical Review},\ }\textbf {\bibinfo
  {volume} {133}},\ \bibinfo {pages} {A1143} (\bibinfo {year}
  {1964})}\BibitemShut {NoStop}%
\bibitem [{\citenamefont {Chaikin}\ and\ \citenamefont
  {Beni}(1976)}]{Chaikin:prb1976}%
  \BibitemOpen
  \bibfield  {author} {\bibinfo {author} {\bibfnamefont {P.~M.}\ \bibnamefont
  {Chaikin}}\ and\ \bibinfo {author} {\bibfnamefont {G.}~\bibnamefont {Beni}},\
  }\href {http://link.aps.org/doi/10.1103/PhysRevB.13.647} {\bibfield
  {journal} {\bibinfo  {journal} {Physical Review B},\ }\textbf {\bibinfo
  {volume} {13}},\ \bibinfo {pages} {647} (\bibinfo {year} {1976})}\BibitemShut
  {NoStop}%
\bibitem [{\citenamefont {P{\'a}lsson}\ and\ \citenamefont
  {Kotliar}(1998)}]{Palsson:prl1998}%
  \BibitemOpen
  \bibfield  {author} {\bibinfo {author} {\bibfnamefont {G.}~\bibnamefont
  {P{\'a}lsson}}\ and\ \bibinfo {author} {\bibfnamefont {G.}~\bibnamefont
  {Kotliar}},\ }\href {http://link.aps.org/doi/10.1103/PhysRevLett.80.4775}
  {\bibfield  {journal} {\bibinfo  {journal} {Physical Review Letters},\
  }\textbf {\bibinfo {volume} {80}},\ \bibinfo {pages} {4775} (\bibinfo {year}
  {1998})}\BibitemShut {NoStop}%
\bibitem [{\citenamefont {Oudovenko}\ \emph {et~al.}(2006)\citenamefont
  {Oudovenko}, \citenamefont {P{\'a}lsson}, \citenamefont {Haule},
  \citenamefont {Kotliar},\ and\ \citenamefont {Savrasov}}]{Oudovenko:prb2006}%
  \BibitemOpen
  \bibfield  {author} {\bibinfo {author} {\bibfnamefont {V.~S.}\ \bibnamefont
  {Oudovenko}}, \bibinfo {author} {\bibfnamefont {G.}~\bibnamefont
  {P{\'a}lsson}}, \bibinfo {author} {\bibfnamefont {K.}~\bibnamefont {Haule}},
  \bibinfo {author} {\bibfnamefont {G.}~\bibnamefont {Kotliar}}, \ and\
  \bibinfo {author} {\bibfnamefont {S.~Y.}\ \bibnamefont {Savrasov}},\ }\href
  {http://link.aps.org/doi/10.1103/PhysRevB.73.035120} {\bibfield  {journal}
  {\bibinfo  {journal} {Physical Review B},\ }\textbf {\bibinfo {volume}
  {73}},\ \bibinfo {pages} {035120} (\bibinfo {year} {2006})}\BibitemShut
  {NoStop}%
\bibitem [{\citenamefont {Tomczak}\ \emph {et~al.}(2010)\citenamefont
  {Tomczak}, \citenamefont {Haule}, \citenamefont {Miyake}, \citenamefont
  {Georges},\ and\ \citenamefont {Kotliar}}]{Tomczak:prb10}%
  \BibitemOpen
  \bibfield  {author} {\bibinfo {author} {\bibfnamefont {J.~M.}\ \bibnamefont
  {Tomczak}}, \bibinfo {author} {\bibfnamefont {K.}~\bibnamefont {Haule}},
  \bibinfo {author} {\bibfnamefont {T.}~\bibnamefont {Miyake}}, \bibinfo
  {author} {\bibfnamefont {A.}~\bibnamefont {Georges}}, \ and\ \bibinfo
  {author} {\bibfnamefont {G.}~\bibnamefont {Kotliar}},\ }\href
  {http://link.aps.org/doi/10.1103/PhysRevB.82.085104} {\bibfield  {journal}
  {\bibinfo  {journal} {Physical Review B},\ }\textbf {\bibinfo {volume}
  {82}},\ \bibinfo {pages} {085104} (\bibinfo {year} {2010})}\BibitemShut
  {NoStop}%
\bibitem [{\citenamefont {Mukerjee}\ and\ \citenamefont
  {Moore}(2007)}]{Mukerjee:APL2007}%
  \BibitemOpen
  \bibfield  {author} {\bibinfo {author} {\bibfnamefont {S.}~\bibnamefont
  {Mukerjee}}\ and\ \bibinfo {author} {\bibfnamefont {J.~E.}\ \bibnamefont
  {Moore}},\ }\href {http://dx.doi.org/10.1063/1.2712775} {\bibfield  {journal}
  {\bibinfo  {journal} {Applied Physics Letters},\ }\textbf {\bibinfo {volume}
  {90}},\ \bibinfo {pages} {112107} (\bibinfo {year} {2007})}\BibitemShut
  {NoStop}%
\bibitem [{\citenamefont {Sakai}\ \emph {et~al.}(2007)\citenamefont {Sakai},
  \citenamefont {Ishii}, \citenamefont {Onose}, \citenamefont {Tomioka},
  \citenamefont {Yotsuhashi}, \citenamefont {Adachi}, \citenamefont {Nagaosa},\
  and\ \citenamefont {Tokura}}]{Sakai:JPSP07}%
  \BibitemOpen
  \bibfield  {author} {\bibinfo {author} {\bibfnamefont {A.}~\bibnamefont
  {Sakai}}, \bibinfo {author} {\bibfnamefont {F.}~\bibnamefont {Ishii}},
  \bibinfo {author} {\bibfnamefont {Y.}~\bibnamefont {Onose}}, \bibinfo
  {author} {\bibfnamefont {Y.}~\bibnamefont {Tomioka}}, \bibinfo {author}
  {\bibfnamefont {S.}~\bibnamefont {Yotsuhashi}}, \bibinfo {author}
  {\bibfnamefont {H.}~\bibnamefont {Adachi}}, \bibinfo {author} {\bibfnamefont
  {N.}~\bibnamefont {Nagaosa}}, \ and\ \bibinfo {author} {\bibfnamefont
  {Y.}~\bibnamefont {Tokura}},\ }\Doi {10.1143/JPSJ.76.093601} {\bibfield
  {journal} {\bibinfo  {journal} {Journal of the Physical Society of Japan},\
  }\textbf {\bibinfo {volume} {76}},\ \bibinfo {pages} {093601} (\bibinfo
  {year} {2007})}\BibitemShut {NoStop}%
\bibitem [{\citenamefont {Terasaki}\ \emph {et~al.}(1997)\citenamefont
  {Terasaki}, \citenamefont {Sasago},\ and\ \citenamefont
  {Uchinokura}}]{Terasaki:prb97}%
  \BibitemOpen
  \bibfield  {author} {\bibinfo {author} {\bibfnamefont {I.}~\bibnamefont
  {Terasaki}}, \bibinfo {author} {\bibfnamefont {Y.}~\bibnamefont {Sasago}}, \
  and\ \bibinfo {author} {\bibfnamefont {K.}~\bibnamefont {Uchinokura}},\
  }\href {http://link.aps.org/doi/10.1103/PhysRevB.56.R12685} {\bibfield
  {journal} {\bibinfo  {journal} {Physical Review B},\ }\textbf {\bibinfo
  {volume} {56}},\ \bibinfo {pages} {R12685} (\bibinfo {year}
  {1997})}\BibitemShut {NoStop}%
\bibitem [{\citenamefont {Haerter}\ \emph {et~al.}(2006)\citenamefont
  {Haerter}, \citenamefont {Peterson},\ and\ \citenamefont
  {Shastry}}]{Haerter:prl06}%
  \BibitemOpen
  \bibfield  {author} {\bibinfo {author} {\bibfnamefont {J.~O.}\ \bibnamefont
  {Haerter}}, \bibinfo {author} {\bibfnamefont {M.~R.}\ \bibnamefont
  {Peterson}}, \ and\ \bibinfo {author} {\bibfnamefont {B.~S.}\ \bibnamefont
  {Shastry}},\ }\href {http://link.aps.org/doi/10.1103/PhysRevLett.97.226402}
  {\bibfield  {journal} {\bibinfo  {journal} {Physical Review Letters},\
  }\textbf {\bibinfo {volume} {97}},\ \bibinfo {pages} {226402} (\bibinfo
  {year} {2006})}\BibitemShut {NoStop}%
\bibitem [{\citenamefont {Okuda}\ \emph {et~al.}(2001)\citenamefont {Okuda},
  \citenamefont {Nakanishi}, \citenamefont {Miyasaka},\ and\ \citenamefont
  {Tokura}}]{Okuda:prb2001}%
  \BibitemOpen
  \bibfield  {author} {\bibinfo {author} {\bibfnamefont {T.}~\bibnamefont
  {Okuda}}, \bibinfo {author} {\bibfnamefont {K.}~\bibnamefont {Nakanishi}},
  \bibinfo {author} {\bibfnamefont {S.}~\bibnamefont {Miyasaka}}, \ and\
  \bibinfo {author} {\bibfnamefont {Y.}~\bibnamefont {Tokura}},\ }\href
  {http://link.aps.org/doi/10.1103/PhysRevB.63.113104} {\bibfield  {journal}
  {\bibinfo  {journal} {Physical Review B},\ }\textbf {\bibinfo {volume}
  {63}},\ \bibinfo {pages} {113104} (\bibinfo {year} {2001})}\BibitemShut
  {NoStop}%
\bibitem [{\citenamefont {Bentien}\ \emph {et~al.}(2007)\citenamefont
  {Bentien}, \citenamefont {Johnsen}, \citenamefont {Madsen}, \citenamefont
  {Iversen},\ and\ \citenamefont {Steglich}}]{Bentien:EPL2007}%
  \BibitemOpen
  \bibfield  {author} {\bibinfo {author} {\bibfnamefont {A.}~\bibnamefont
  {Bentien}}, \bibinfo {author} {\bibfnamefont {S.}~\bibnamefont {Johnsen}},
  \bibinfo {author} {\bibfnamefont {G.~K.~H.}\ \bibnamefont {Madsen}}, \bibinfo
  {author} {\bibfnamefont {B.~B.}\ \bibnamefont {Iversen}}, \ and\ \bibinfo
  {author} {\bibfnamefont {F.}~\bibnamefont {Steglich}},\ }\href
  {http://stacks.iop.org/0295-5075/80/i=1/a=17008} {\bibfield  {journal}
  {\bibinfo  {journal} {EPL (Europhysics Letters)},\ }\textbf {\bibinfo
  {volume} {80}},\ \bibinfo {pages} {17008} (\bibinfo {year}
  {2007})}\BibitemShut {NoStop}%
\bibitem [{\citenamefont {Venkatasubramanian}\ \emph
  {et~al.}(2001)\citenamefont {Venkatasubramanian}, \citenamefont {Siivola},
  \citenamefont {Colpitts},\ and\ \citenamefont
  {O'Quinn}}]{Venkatasubramanian:na2001}%
  \BibitemOpen
  \bibfield  {author} {\bibinfo {author} {\bibfnamefont {R.}~\bibnamefont
  {Venkatasubramanian}}, \bibinfo {author} {\bibfnamefont {E.}~\bibnamefont
  {Siivola}}, \bibinfo {author} {\bibfnamefont {T.}~\bibnamefont {Colpitts}}, \
  and\ \bibinfo {author} {\bibfnamefont {B.}~\bibnamefont {O'Quinn}},\ }\href
  {http://dx.doi.org/10.1038/35098012} {\bibfield  {journal} {\bibinfo
  {journal} {Nature},\ }\textbf {\bibinfo {volume} {413}},\ \bibinfo {pages}
  {597} (\bibinfo {year} {2001})}\BibitemShut {NoStop}%
\bibitem [{\citenamefont {Ravichandran}\ \emph {et~al.}(2012)\citenamefont
  {Ravichandran}, \citenamefont {Yadav}, \citenamefont {Siemons}, \citenamefont
  {McGuire}, \citenamefont {Wu}, \citenamefont {Vailionis}, \citenamefont
  {Majumdar},\ and\ \citenamefont {Ramesh}}]{Ravichandran:prb12}%
  \BibitemOpen
  \bibfield  {author} {\bibinfo {author} {\bibfnamefont {J.}~\bibnamefont
  {Ravichandran}}, \bibinfo {author} {\bibfnamefont {A.~K.}\ \bibnamefont
  {Yadav}}, \bibinfo {author} {\bibfnamefont {W.}~\bibnamefont {Siemons}},
  \bibinfo {author} {\bibfnamefont {M.~A.}\ \bibnamefont {McGuire}}, \bibinfo
  {author} {\bibfnamefont {V.}~\bibnamefont {Wu}}, \bibinfo {author}
  {\bibfnamefont {A.}~\bibnamefont {Vailionis}}, \bibinfo {author}
  {\bibfnamefont {A.}~\bibnamefont {Majumdar}}, \ and\ \bibinfo {author}
  {\bibfnamefont {R.}~\bibnamefont {Ramesh}},\ }\Doi
  {10.1103/PhysRevB.85.085112} {\bibfield  {journal} {\bibinfo  {journal}
  {Phys. Rev. B},\ }\textbf {\bibinfo {volume} {85}},\ \bibinfo {pages}
  {085112} (\bibinfo {year} {2012})}\BibitemShut {NoStop}%
\bibitem [{\citenamefont {Koshibae}\ \emph {et~al.}(2000)\citenamefont
  {Koshibae}, \citenamefont {Tsutsui},\ and\ \citenamefont
  {Maekawa}}]{Koshibae:prb2000}%
  \BibitemOpen
  \bibfield  {author} {\bibinfo {author} {\bibfnamefont {W.}~\bibnamefont
  {Koshibae}}, \bibinfo {author} {\bibfnamefont {K.}~\bibnamefont {Tsutsui}}, \
  and\ \bibinfo {author} {\bibfnamefont {S.}~\bibnamefont {Maekawa}},\ }\href
  {http://link.aps.org/doi/10.1103/PhysRevB.62.6869} {\bibfield  {journal}
  {\bibinfo  {journal} {Physical Review B},\ }\textbf {\bibinfo {volume}
  {62}},\ \bibinfo {pages} {6869} (\bibinfo {year} {2000})}\BibitemShut
  {NoStop}%
\bibitem [{\citenamefont {Mukerjee}(2005)}]{Mukerjee:prb2005}%
  \BibitemOpen
  \bibfield  {author} {\bibinfo {author} {\bibfnamefont {S.}~\bibnamefont
  {Mukerjee}},\ }\href {http://link.aps.org/doi/10.1103/PhysRevB.72.195109}
  {\bibfield  {journal} {\bibinfo  {journal} {Physical Review B},\ }\textbf
  {\bibinfo {volume} {72}},\ \bibinfo {pages} {195109} (\bibinfo {year}
  {2005})}\BibitemShut {NoStop}%
\bibitem [{\citenamefont {Uchida}\ \emph {et~al.}(2011)\citenamefont {Uchida},
  \citenamefont {Oishi}, \citenamefont {Matsuo}, \citenamefont {Koshibae},
  \citenamefont {Onose}, \citenamefont {Mori}, \citenamefont {Fujioka},
  \citenamefont {Miyasaka}, \citenamefont {Maekawa},\ and\ \citenamefont
  {Tokura}}]{Uchida:prb11}%
  \BibitemOpen
  \bibfield  {author} {\bibinfo {author} {\bibfnamefont {M.}~\bibnamefont
  {Uchida}}, \bibinfo {author} {\bibfnamefont {K.}~\bibnamefont {Oishi}},
  \bibinfo {author} {\bibfnamefont {M.}~\bibnamefont {Matsuo}}, \bibinfo
  {author} {\bibfnamefont {W.}~\bibnamefont {Koshibae}}, \bibinfo {author}
  {\bibfnamefont {Y.}~\bibnamefont {Onose}}, \bibinfo {author} {\bibfnamefont
  {M.}~\bibnamefont {Mori}}, \bibinfo {author} {\bibfnamefont {J.}~\bibnamefont
  {Fujioka}}, \bibinfo {author} {\bibfnamefont {S.}~\bibnamefont {Miyasaka}},
  \bibinfo {author} {\bibfnamefont {S.}~\bibnamefont {Maekawa}}, \ and\
  \bibinfo {author} {\bibfnamefont {Y.}~\bibnamefont {Tokura}},\ }\href
  {http://link.aps.org/doi/10.1103/PhysRevB.83.165127} {\bibfield  {journal}
  {\bibinfo  {journal} {Physical Review B},\ }\textbf {\bibinfo {volume}
  {83}},\ \bibinfo {pages} {165127} (\bibinfo {year} {2011})}\BibitemShut
  {NoStop}%
\bibitem [{\citenamefont {Oguri}\ and\ \citenamefont
  {Maekawa}(1990)}]{Oguri:prb1990}%
  \BibitemOpen
  \bibfield  {author} {\bibinfo {author} {\bibfnamefont {A.}~\bibnamefont
  {Oguri}}\ and\ \bibinfo {author} {\bibfnamefont {S.}~\bibnamefont
  {Maekawa}},\ }\href {http://link.aps.org/doi/10.1103/PhysRevB.41.6977}
  {\bibfield  {journal} {\bibinfo  {journal} {Physical Review B},\ }\textbf
  {\bibinfo {volume} {41}},\ \bibinfo {pages} {6977} (\bibinfo {year}
  {1990})}\BibitemShut {NoStop}%
\bibitem [{\citenamefont {Paul}\ and\ \citenamefont
  {Kotliar}(2003)}]{Paul:prb2003}%
  \BibitemOpen
  \bibfield  {author} {\bibinfo {author} {\bibfnamefont {I.}~\bibnamefont
  {Paul}}\ and\ \bibinfo {author} {\bibfnamefont {G.}~\bibnamefont {Kotliar}},\
  }\href {http://link.aps.org/doi/10.1103/PhysRevB.67.115131} {\bibfield
  {journal} {\bibinfo  {journal} {Physical Review B},\ }\textbf {\bibinfo
  {volume} {67}},\ \bibinfo {pages} {115131} (\bibinfo {year}
  {2003})}\BibitemShut {NoStop}%
\bibitem [{\citenamefont {Tokura}\ \emph {et~al.}(1994)\citenamefont {Tokura},
  \citenamefont {Urushibara}, \citenamefont {Moritomo}, \citenamefont {Arima},
  \citenamefont {Asamitsu}, \citenamefont {Kido},\ and\ \citenamefont
  {Furukawa}}]{Tokura:JPSJ94}%
  \BibitemOpen
  \bibfield  {author} {\bibinfo {author} {\bibfnamefont {Y.}~\bibnamefont
  {Tokura}}, \bibinfo {author} {\bibfnamefont {A.}~\bibnamefont {Urushibara}},
  \bibinfo {author} {\bibfnamefont {Y.}~\bibnamefont {Moritomo}}, \bibinfo
  {author} {\bibfnamefont {T.}~\bibnamefont {Arima}}, \bibinfo {author}
  {\bibfnamefont {A.}~\bibnamefont {Asamitsu}}, \bibinfo {author}
  {\bibfnamefont {G.}~\bibnamefont {Kido}}, \ and\ \bibinfo {author}
  {\bibfnamefont {N.}~\bibnamefont {Furukawa}},\ }\Doi {10.1143/JPSJ.63.3931}
  {\bibfield  {journal} {\bibinfo  {journal} {Journal of the Physical Society
  of Japan},\ }\textbf {\bibinfo {volume} {63}},\ \bibinfo {pages} {3931}
  (\bibinfo {year} {1994})}\BibitemShut {NoStop}%
\bibitem [{\citenamefont {Maeno}\ \emph {et~al.}(1994)\citenamefont {Maeno},
  \citenamefont {Hashimoto}, \citenamefont {Yoshida}, \citenamefont
  {Nishizaki}, \citenamefont {Fujita}, \citenamefont {Bednorz},\ and\
  \citenamefont {Lichtenberg}}]{Maeno:na1994}%
  \BibitemOpen
  \bibfield  {author} {\bibinfo {author} {\bibfnamefont {Y.}~\bibnamefont
  {Maeno}}, \bibinfo {author} {\bibfnamefont {H.}~\bibnamefont {Hashimoto}},
  \bibinfo {author} {\bibfnamefont {K.}~\bibnamefont {Yoshida}}, \bibinfo
  {author} {\bibfnamefont {S.}~\bibnamefont {Nishizaki}}, \bibinfo {author}
  {\bibfnamefont {T.}~\bibnamefont {Fujita}}, \bibinfo {author} {\bibfnamefont
  {J.~G.}\ \bibnamefont {Bednorz}}, \ and\ \bibinfo {author} {\bibfnamefont
  {F.}~\bibnamefont {Lichtenberg}},\ }\href
  {http://dx.doi.org/10.1038/372532a0} {\bibfield  {journal} {\bibinfo
  {journal} {Nature},\ }\textbf {\bibinfo {volume} {372}},\ \bibinfo {pages}
  {532} (\bibinfo {year} {1994})}\BibitemShut {NoStop}%
\bibitem [{\citenamefont {Kamihara}\ \emph {et~al.}(2008)\citenamefont
  {Kamihara}, \citenamefont {Watanabe}, \citenamefont {Hirano},\ and\
  \citenamefont {Hosono}}]{Kamihara:JACS2008}%
  \BibitemOpen
  \bibfield  {author} {\bibinfo {author} {\bibfnamefont {Y.}~\bibnamefont
  {Kamihara}}, \bibinfo {author} {\bibfnamefont {T.}~\bibnamefont {Watanabe}},
  \bibinfo {author} {\bibfnamefont {M.}~\bibnamefont {Hirano}}, \ and\ \bibinfo
  {author} {\bibfnamefont {H.}~\bibnamefont {Hosono}},\ }\bibfield  {booktitle}
  {\emph {\bibinfo {booktitle} {Journal of the American Chemical Society}},\
  }\Doi {10.1021/ja800073m} {\bibfield  {journal} {\bibinfo  {journal} {Journal
  of the American Chemical Society},\ }\textbf {\bibinfo {volume} {130}},\
  \bibinfo {pages} {3296} (\bibinfo {year} {2008})}\BibitemShut {NoStop}%
\bibitem [{\citenamefont {Georges}\ \emph {et~al.}(1996)\citenamefont
  {Georges}, \citenamefont {Kotliar}, \citenamefont {Krauth},\ and\
  \citenamefont {Rozenberg}}]{Georges:rmp1996}%
  \BibitemOpen
  \bibfield  {author} {\bibinfo {author} {\bibfnamefont {A.}~\bibnamefont
  {Georges}}, \bibinfo {author} {\bibfnamefont {G.}~\bibnamefont {Kotliar}},
  \bibinfo {author} {\bibfnamefont {W.}~\bibnamefont {Krauth}}, \ and\ \bibinfo
  {author} {\bibfnamefont {M.~J.}\ \bibnamefont {Rozenberg}},\ }\href
  {http://link.aps.org/doi/10.1103/RevModPhys.68.13} {\bibfield  {journal}
  {\bibinfo  {journal} {Reviews of Modern Physics},\ }\textbf {\bibinfo
  {volume} {68}},\ \bibinfo {pages} {13} (\bibinfo {year} {1996})}\BibitemShut
  {NoStop}%
\bibitem [{\citenamefont {Metzner}\ and\ \citenamefont
  {Vollhardt}(1989)}]{Metzner:prl1989}%
  \BibitemOpen
  \bibfield  {author} {\bibinfo {author} {\bibfnamefont {W.}~\bibnamefont
  {Metzner}}\ and\ \bibinfo {author} {\bibfnamefont {D.}~\bibnamefont
  {Vollhardt}},\ }\href {http://link.aps.org/doi/10.1103/PhysRevLett.62.324}
  {\bibfield  {journal} {\bibinfo  {journal} {Physical Review Letters},\
  }\textbf {\bibinfo {volume} {62}},\ \bibinfo {pages} {324} (\bibinfo {year}
  {1989})}\BibitemShut {NoStop}%
\bibitem [{\citenamefont {Caffarel}\ and\ \citenamefont
  {Krauth}(1994)}]{Caffarel:prl1994}%
  \BibitemOpen
  \bibfield  {author} {\bibinfo {author} {\bibfnamefont {M.}~\bibnamefont
  {Caffarel}}\ and\ \bibinfo {author} {\bibfnamefont {W.}~\bibnamefont
  {Krauth}},\ }\href {http://link.aps.org/doi/10.1103/PhysRevLett.72.1545}
  {\bibfield  {journal} {\bibinfo  {journal} {Physical Review Letters},\
  }\textbf {\bibinfo {volume} {72}},\ \bibinfo {pages} {1545} (\bibinfo {year}
  {1994})}\BibitemShut {NoStop}%
\bibitem [{\citenamefont {Werner}\ \emph {et~al.}(2006)\citenamefont {Werner},
  \citenamefont {Comanac}, \citenamefont {de'Medici}, \citenamefont {Troyer},\
  and\ \citenamefont {Millis}}]{Werner:prl2006}%
  \BibitemOpen
  \bibfield  {author} {\bibinfo {author} {\bibfnamefont {P.}~\bibnamefont
  {Werner}}, \bibinfo {author} {\bibfnamefont {A.}~\bibnamefont {Comanac}},
  \bibinfo {author} {\bibfnamefont {L.}~\bibnamefont {de'Medici}}, \bibinfo
  {author} {\bibfnamefont {M.}~\bibnamefont {Troyer}}, \ and\ \bibinfo {author}
  {\bibfnamefont {A.~J.}\ \bibnamefont {Millis}},\ }\href
  {http://link.aps.org/doi/10.1103/PhysRevLett.97.076405} {\bibfield  {journal}
  {\bibinfo  {journal} {Physical Review Letters},\ }\textbf {\bibinfo {volume}
  {97}},\ \bibinfo {pages} {076405} (\bibinfo {year} {2006})}\BibitemShut
  {NoStop}%
\bibitem [{\citenamefont {Werner}\ and\ \citenamefont
  {Millis}(2006)}]{Werner:prb2006}%
  \BibitemOpen
  \bibfield  {author} {\bibinfo {author} {\bibfnamefont {P.}~\bibnamefont
  {Werner}}\ and\ \bibinfo {author} {\bibfnamefont {A.~J.}\ \bibnamefont
  {Millis}},\ }\href {http://link.aps.org/doi/10.1103/PhysRevB.74.155107}
  {\bibfield  {journal} {\bibinfo  {journal} {Physical Review B},\ }\textbf
  {\bibinfo {volume} {74}},\ \bibinfo {pages} {155107} (\bibinfo {year}
  {2006})}\BibitemShut {NoStop}%
\bibitem [{\citenamefont {Gull}\ \emph {et~al.}(2011)\citenamefont {Gull},
  \citenamefont {Millis}, \citenamefont {Lichtenstein}, \citenamefont
  {Rubtsov}, \citenamefont {Troyer},\ and\ \citenamefont
  {Werner}}]{Gull:rmp2011}%
  \BibitemOpen
  \bibfield  {author} {\bibinfo {author} {\bibfnamefont {E.}~\bibnamefont
  {Gull}}, \bibinfo {author} {\bibfnamefont {A.~J.}\ \bibnamefont {Millis}},
  \bibinfo {author} {\bibfnamefont {A.~I.}\ \bibnamefont {Lichtenstein}},
  \bibinfo {author} {\bibfnamefont {A.~N.}\ \bibnamefont {Rubtsov}}, \bibinfo
  {author} {\bibfnamefont {M.}~\bibnamefont {Troyer}}, \ and\ \bibinfo {author}
  {\bibfnamefont {P.}~\bibnamefont {Werner}},\ }\href
  {http://link.aps.org/doi/10.1103/RevModPhys.83.349} {\bibfield  {journal}
  {\bibinfo  {journal} {Reviews of Modern Physics},\ }\textbf {\bibinfo
  {volume} {83}},\ \bibinfo {pages} {349} (\bibinfo {year} {2011})}\BibitemShut
  {NoStop}%
\bibitem [{\citenamefont {R\"uegg}\ \emph {et~al.}(2013)\citenamefont
  {R\"uegg}, \citenamefont {Gull}, \citenamefont {Fiete},\ and\ \citenamefont
  {Millis}}]{Ruegg:prb13}%
  \BibitemOpen
  \bibfield  {author} {\bibinfo {author} {\bibfnamefont {A.}~\bibnamefont
  {R\"uegg}}, \bibinfo {author} {\bibfnamefont {E.}~\bibnamefont {Gull}},
  \bibinfo {author} {\bibfnamefont {G.~A.}\ \bibnamefont {Fiete}}, \ and\
  \bibinfo {author} {\bibfnamefont {A.~J.}\ \bibnamefont {Millis}},\ }\Doi
  {10.1103/PhysRevB.87.075124} {\bibfield  {journal} {\bibinfo  {journal}
  {Phys. Rev. B},\ }\textbf {\bibinfo {volume} {87}},\ \bibinfo {pages}
  {075124} (\bibinfo {year} {2013})}\BibitemShut {NoStop}%
\bibitem [{\citenamefont {Georges}\ and\ \citenamefont
  {Kotliar}(1992)}]{Georges:prb1992}%
  \BibitemOpen
  \bibfield  {author} {\bibinfo {author} {\bibfnamefont {A.}~\bibnamefont
  {Georges}}\ and\ \bibinfo {author} {\bibfnamefont {G.}~\bibnamefont
  {Kotliar}},\ }\href {http://link.aps.org/doi/10.1103/PhysRevB.45.6479}
  {\bibfield  {journal} {\bibinfo  {journal} {Physical Review B},\ }\textbf
  {\bibinfo {volume} {45}},\ \bibinfo {pages} {6479} (\bibinfo {year}
  {1992})}\BibitemShut {NoStop}%
\bibitem [{\citenamefont {Rozenberg}\ \emph {et~al.}(1994)\citenamefont
  {Rozenberg}, \citenamefont {Kotliar},\ and\ \citenamefont
  {Zhang}}]{Rozenberg:prb1994}%
  \BibitemOpen
  \bibfield  {author} {\bibinfo {author} {\bibfnamefont {M.~J.}\ \bibnamefont
  {Rozenberg}}, \bibinfo {author} {\bibfnamefont {G.}~\bibnamefont {Kotliar}},
  \ and\ \bibinfo {author} {\bibfnamefont {X.~Y.}\ \bibnamefont {Zhang}},\
  }\href {http://link.aps.org/doi/10.1103/PhysRevB.49.10181} {\bibfield
  {journal} {\bibinfo  {journal} {Physical Review B},\ }\textbf {\bibinfo
  {volume} {49}},\ \bibinfo {pages} {10181} (\bibinfo {year}
  {1994})}\BibitemShut {NoStop}%
\bibitem [{\citenamefont {Rozenberg}\ \emph {et~al.}(1995)\citenamefont
  {Rozenberg}, \citenamefont {Kotliar}, \citenamefont {Kajueter}, \citenamefont
  {Thomas}, \citenamefont {Rapkine}, \citenamefont {Honig},\ and\ \citenamefont
  {Metcalf}}]{Rozenberg:prl1995}%
  \BibitemOpen
  \bibfield  {author} {\bibinfo {author} {\bibfnamefont {M.~J.}\ \bibnamefont
  {Rozenberg}}, \bibinfo {author} {\bibfnamefont {G.}~\bibnamefont {Kotliar}},
  \bibinfo {author} {\bibfnamefont {H.}~\bibnamefont {Kajueter}}, \bibinfo
  {author} {\bibfnamefont {G.~A.}\ \bibnamefont {Thomas}}, \bibinfo {author}
  {\bibfnamefont {D.~H.}\ \bibnamefont {Rapkine}}, \bibinfo {author}
  {\bibfnamefont {J.~M.}\ \bibnamefont {Honig}}, \ and\ \bibinfo {author}
  {\bibfnamefont {P.}~\bibnamefont {Metcalf}},\ }\href
  {http://link.aps.org/doi/10.1103/PhysRevLett.75.105} {\bibfield  {journal}
  {\bibinfo  {journal} {Physical Review Letters},\ }\textbf {\bibinfo {volume}
  {75}},\ \bibinfo {pages} {105} (\bibinfo {year} {1995})}\BibitemShut
  {NoStop}%
\bibitem [{\citenamefont {Matsuo}\ \emph {et~al.}(2011)\citenamefont {Matsuo},
  \citenamefont {Okamoto}, \citenamefont {Koshibae}, \citenamefont {Mori},\
  and\ \citenamefont {Maekawa}}]{Matsuo:prb11}%
  \BibitemOpen
  \bibfield  {author} {\bibinfo {author} {\bibfnamefont {M.}~\bibnamefont
  {Matsuo}}, \bibinfo {author} {\bibfnamefont {S.}~\bibnamefont {Okamoto}},
  \bibinfo {author} {\bibfnamefont {W.}~\bibnamefont {Koshibae}}, \bibinfo
  {author} {\bibfnamefont {M.}~\bibnamefont {Mori}}, \ and\ \bibinfo {author}
  {\bibfnamefont {S.}~\bibnamefont {Maekawa}},\ }\Doi
  {10.1103/PhysRevB.84.153107} {\bibfield  {journal} {\bibinfo  {journal}
  {Phys. Rev. B},\ }\textbf {\bibinfo {volume} {84}},\ \bibinfo {pages}
  {153107} (\bibinfo {year} {2011})}\BibitemShut {NoStop}%
\bibitem [{\citenamefont {Barman}\ and\ \citenamefont
  {Vidhyadhiraja}(2011)}]{H.BARMAN:ijpb2011}%
  \BibitemOpen
  \bibfield  {author} {\bibinfo {author} {\bibfnamefont {H.}~\bibnamefont
  {Barman}}\ and\ \bibinfo {author} {\bibfnamefont {N.~S.}\ \bibnamefont
  {Vidhyadhiraja}},\ }\href
  {http://www.worldscientific.com/doi/abs/10.1142/S0217979211100977} {\bibfield
   {journal} {\bibinfo  {journal} {International Journal of Modern Physics B},\
  }\textbf {\bibinfo {volume} {25}},\ \bibinfo {pages} {2461} (\bibinfo {year}
  {2011})}\BibitemShut {NoStop}%
\bibitem [{\citenamefont {Kajueter}\ and\ \citenamefont
  {Kotliar}(1996)}]{Kajueter:prl1996}%
  \BibitemOpen
  \bibfield  {author} {\bibinfo {author} {\bibfnamefont {H.}~\bibnamefont
  {Kajueter}}\ and\ \bibinfo {author} {\bibfnamefont {G.}~\bibnamefont
  {Kotliar}},\ }\href {http://link.aps.org/doi/10.1103/PhysRevLett.77.131}
  {\bibfield  {journal} {\bibinfo  {journal} {Physical Review Letters},\
  }\textbf {\bibinfo {volume} {77}},\ \bibinfo {pages} {131} (\bibinfo {year}
  {1996})}\BibitemShut {NoStop}%
\bibitem [{\citenamefont {Langreth}(1966)}]{Langreth:pr1966}%
  \BibitemOpen
  \bibfield  {author} {\bibinfo {author} {\bibfnamefont {D.~C.}\ \bibnamefont
  {Langreth}},\ }\href {http://link.aps.org/doi/10.1103/PhysRev.150.516}
  {\bibfield  {journal} {\bibinfo  {journal} {Physical Review},\ }\textbf
  {\bibinfo {volume} {150}},\ \bibinfo {pages} {516} (\bibinfo {year}
  {1966})}\BibitemShut {NoStop}%
\bibitem [{\citenamefont {Luttinger}\ and\ \citenamefont
  {Ward}(1960)}]{Luttinger:pr1960}%
  \BibitemOpen
  \bibfield  {author} {\bibinfo {author} {\bibfnamefont {J.~M.}\ \bibnamefont
  {Luttinger}}\ and\ \bibinfo {author} {\bibfnamefont {J.~C.}\ \bibnamefont
  {Ward}},\ }\href {http://link.aps.org/doi/10.1103/PhysRev.118.1417}
  {\bibfield  {journal} {\bibinfo  {journal} {Physical Review},\ }\textbf
  {\bibinfo {volume} {118}},\ \bibinfo {pages} {1417} (\bibinfo {year}
  {1960})}\BibitemShut {NoStop}%
\bibitem [{\citenamefont {Arsenault}\ \emph {et~al.}(2012)\citenamefont
  {Arsenault}, \citenamefont {S{\'e}mon},\ and\ \citenamefont
  {Tremblay}}]{Arsenault:prb2012}%
  \BibitemOpen
  \bibfield  {author} {\bibinfo {author} {\bibfnamefont {L.-F.}\ \bibnamefont
  {Arsenault}}, \bibinfo {author} {\bibfnamefont {P.}~\bibnamefont
  {S{\'e}mon}}, \ and\ \bibinfo {author} {\bibfnamefont {A.~M.~S.}\
  \bibnamefont {Tremblay}},\ }\href
  {http://link.aps.org/doi/10.1103/PhysRevB.86.085133} {\bibfield  {journal}
  {\bibinfo  {journal} {Physical Review B},\ }\textbf {\bibinfo {volume}
  {86}},\ \bibinfo {pages} {085133} (\bibinfo {year} {2012})}\BibitemShut
  {NoStop}%
\bibitem [{\citenamefont {Kotliar}\ and\ \citenamefont
  {Kajueter}(1996)}]{Kotliar:prb1996}%
  \BibitemOpen
  \bibfield  {author} {\bibinfo {author} {\bibfnamefont {G.}~\bibnamefont
  {Kotliar}}\ and\ \bibinfo {author} {\bibfnamefont {H.}~\bibnamefont
  {Kajueter}},\ }\href {http://link.aps.org/doi/10.1103/PhysRevB.54.R14221}
  {\bibfield  {journal} {\bibinfo  {journal} {Physical Review B},\ }\textbf
  {\bibinfo {volume} {54}},\ \bibinfo {pages} {R14221} (\bibinfo {year}
  {1996})}\BibitemShut {NoStop}%
\bibitem [{\citenamefont {Yeyati}\ \emph {et~al.}(1999)\citenamefont {Yeyati},
  \citenamefont {Flores},\ and\ \citenamefont
  {Mart{\'\i}n-Rodero}}]{Yeyati:prl1999}%
  \BibitemOpen
  \bibfield  {author} {\bibinfo {author} {\bibfnamefont {A.~L.}\ \bibnamefont
  {Yeyati}}, \bibinfo {author} {\bibfnamefont {F.}~\bibnamefont {Flores}}, \
  and\ \bibinfo {author} {\bibfnamefont {A.}~\bibnamefont
  {Mart{\'\i}n-Rodero}},\ }\href
  {http://link.aps.org/doi/10.1103/PhysRevLett.83.600} {\bibfield  {journal}
  {\bibinfo  {journal} {Physical Review Letters},\ }\textbf {\bibinfo {volume}
  {83}},\ \bibinfo {pages} {600} (\bibinfo {year} {1999})}\BibitemShut
  {NoStop}%
\bibitem [{\citenamefont {Kajueter}(1996)}]{Kajueter:thesis}%
  \BibitemOpen
  \bibfield  {author} {\bibinfo {author} {\bibfnamefont {H.}~\bibnamefont
  {Kajueter}},\ }\href@noop {} {\emph {\bibinfo {title} {Ph.D. thesis, Rutgers
  University Graduate School}}}\ (\bibinfo  {publisher} {New Brunswick, NJ},\
  \bibinfo {year} {1996})\BibitemShut {NoStop}%
\bibitem [{\citenamefont {Mahan}(2005)}]{mahan:book}%
  \BibitemOpen
  \bibfield  {author} {\bibinfo {author} {\bibfnamefont {G.}~\bibnamefont
  {Mahan}},\ }\href@noop {} {\emph {\bibinfo {title} {Many-Body Particles}}}\
  (\bibinfo  {publisher} {Kluwer Academic Plenum Publishers},\ \bibinfo {year}
  {2005})\BibitemShut {NoStop}%
\bibitem [{\citenamefont {Arsenault}\ and\ \citenamefont
  {Tremblay}(2013)}]{Arsenault:arxiv2013}%
  \BibitemOpen
  \bibfield  {author} {\bibinfo {author} {\bibfnamefont {L.-F.}\ \bibnamefont
  {Arsenault}}\ and\ \bibinfo {author} {\bibfnamefont {A.~M.~S.}\ \bibnamefont
  {Tremblay}},\ }\href@noop {} {\bibfield  {journal} {\bibinfo  {journal}
  {arXiv:1305.6999}} (\bibinfo {year} {2013})}\BibitemShut {NoStop}%
\bibitem [{\citenamefont {Arsenault}\ \emph {et~al.}(2013)\citenamefont
  {Arsenault}, \citenamefont {Shastry}, \citenamefont {S{\'e}mon},\ and\
  \citenamefont {Tremblay}}]{Arsenault:prb13}%
  \BibitemOpen
  \bibfield  {author} {\bibinfo {author} {\bibfnamefont {L.-F.}\ \bibnamefont
  {Arsenault}}, \bibinfo {author} {\bibfnamefont {B.~S.}\ \bibnamefont
  {Shastry}}, \bibinfo {author} {\bibfnamefont {P.}~\bibnamefont {S{\'e}mon}},
  \ and\ \bibinfo {author} {\bibfnamefont {A.~M.~S.}\ \bibnamefont
  {Tremblay}},\ }\href {http://link.aps.org/doi/10.1103/PhysRevB.87.035126}
  {\bibfield  {journal} {\bibinfo  {journal} {Physical Review B},\ }\textbf
  {\bibinfo {volume} {87}},\ \bibinfo {pages} {035126} (\bibinfo {year}
  {2013})}\BibitemShut {NoStop}%
\bibitem [{\citenamefont {Chakraborty}\ \emph {et~al.}(2010)\citenamefont
  {Chakraborty}, \citenamefont {Galanakis},\ and\ \citenamefont
  {Phillips}}]{Chakraborty:prb10}%
  \BibitemOpen
  \bibfield  {author} {\bibinfo {author} {\bibfnamefont {S.}~\bibnamefont
  {Chakraborty}}, \bibinfo {author} {\bibfnamefont {D.}~\bibnamefont
  {Galanakis}}, \ and\ \bibinfo {author} {\bibfnamefont {P.}~\bibnamefont
  {Phillips}},\ }\href {http://link.aps.org/doi/10.1103/PhysRevB.82.214503}
  {\bibfield  {journal} {\bibinfo  {journal} {Physical Review B},\ }\textbf
  {\bibinfo {volume} {82}},\ \bibinfo {pages} {214503} (\bibinfo {year}
  {2010})}\BibitemShut {NoStop}%
\bibitem [{\citenamefont {Xu}\ \emph {et~al.}(2013)\citenamefont {Xu},
  \citenamefont {Haule},\ and\ \citenamefont {Kotliar}}]{Xu:prl2013}%
  \BibitemOpen
  \bibfield  {author} {\bibinfo {author} {\bibfnamefont {W.}~\bibnamefont
  {Xu}}, \bibinfo {author} {\bibfnamefont {K.}~\bibnamefont {Haule}}, \ and\
  \bibinfo {author} {\bibfnamefont {G.}~\bibnamefont {Kotliar}},\ }\href
  {http://link.aps.org/doi/10.1103/PhysRevLett.111.036401} {\bibfield
  {journal} {\bibinfo  {journal} {Physical Review Letters},\ }\textbf {\bibinfo
  {volume} {111}},\ \bibinfo {pages} {036401} (\bibinfo {year}
  {2013})}\BibitemShut {NoStop}%
\bibitem [{\citenamefont {Hong}\ \emph {et~al.}(2013)\citenamefont {Hong},
  \citenamefont {Ghaemi}, \citenamefont {Moore},\ and\ \citenamefont
  {Phillips}}]{Hong:prb13}%
  \BibitemOpen
  \bibfield  {author} {\bibinfo {author} {\bibfnamefont {S.}~\bibnamefont
  {Hong}}, \bibinfo {author} {\bibfnamefont {P.}~\bibnamefont {Ghaemi}},
  \bibinfo {author} {\bibfnamefont {J.~E.}\ \bibnamefont {Moore}}, \ and\
  \bibinfo {author} {\bibfnamefont {P.~W.}\ \bibnamefont {Phillips}},\ }\href
  {http://link.aps.org/doi/10.1103/PhysRevB.88.075118} {\bibfield  {journal}
  {\bibinfo  {journal} {Physical Review B},\ }\textbf {\bibinfo {volume}
  {88}},\ \bibinfo {pages} {075118} (\bibinfo {year} {2013})}\BibitemShut
  {NoStop}%
\bibitem [{\citenamefont {Sekino}\ \emph {et~al.}(2013)\citenamefont {Sekino},
  \citenamefont {Okamoto}, \citenamefont {Koshibae}, \citenamefont {Mori},\
  and\ \citenamefont {Maekawa}}]{Sekino:arxiv1308.0416}%
  \BibitemOpen
  \bibfield  {author} {\bibinfo {author} {\bibfnamefont {M.}~\bibnamefont
  {Sekino}}, \bibinfo {author} {\bibfnamefont {S.}~\bibnamefont {Okamoto}},
  \bibinfo {author} {\bibfnamefont {W.}~\bibnamefont {Koshibae}}, \bibinfo
  {author} {\bibfnamefont {M.}~\bibnamefont {Mori}}, \ and\ \bibinfo {author}
  {\bibfnamefont {S.}~\bibnamefont {Maekawa}},\ }\href@noop {} {\bibfield
  {journal} {\bibinfo  {journal} {arXiv:1308.0416}} (\bibinfo {year}
  {2013})}\BibitemShut {NoStop}%
\bibitem [{\citenamefont {Merino}\ and\ \citenamefont
  {McKenzie}(2000)}]{Merino:prb2000}%
  \BibitemOpen
  \bibfield  {author} {\bibinfo {author} {\bibfnamefont {J.}~\bibnamefont
  {Merino}}\ and\ \bibinfo {author} {\bibfnamefont {R.~H.}\ \bibnamefont
  {McKenzie}},\ }\href {http://link.aps.org/doi/10.1103/PhysRevB.61.7996}
  {\bibfield  {journal} {\bibinfo  {journal} {Physical Review B},\ }\textbf
  {\bibinfo {volume} {61}},\ \bibinfo {pages} {7996} (\bibinfo {year}
  {2000})}\BibitemShut {NoStop}%
\bibitem [{\citenamefont {Deng}\ \emph {et~al.}(2013)\citenamefont {Deng},
  \citenamefont {Mravlje}, \citenamefont {{\v Z}itko}, \citenamefont {Ferrero},
  \citenamefont {Kotliar},\ and\ \citenamefont {Georges}}]{Deng:prl13}%
  \BibitemOpen
  \bibfield  {author} {\bibinfo {author} {\bibfnamefont {X.}~\bibnamefont
  {Deng}}, \bibinfo {author} {\bibfnamefont {J.}~\bibnamefont {Mravlje}},
  \bibinfo {author} {\bibfnamefont {R.}~\bibnamefont {{\v Z}itko}}, \bibinfo
  {author} {\bibfnamefont {M.}~\bibnamefont {Ferrero}}, \bibinfo {author}
  {\bibfnamefont {G.}~\bibnamefont {Kotliar}}, \ and\ \bibinfo {author}
  {\bibfnamefont {A.}~\bibnamefont {Georges}},\ }\href
  {http://link.aps.org/doi/10.1103/PhysRevLett.110.086401} {\bibfield
  {journal} {\bibinfo  {journal} {Physical Review Letters},\ }\textbf {\bibinfo
  {volume} {110}},\ \bibinfo {pages} {086401} (\bibinfo {year}
  {2013})}\BibitemShut {NoStop}%
\bibitem [{\citenamefont {Zlati\ifmmode~\acute{c}\else \'{c}\fi{}}\ and\
  \citenamefont {Freericks}(2012)}]{Zlatic:prl12}%
  \BibitemOpen
  \bibfield  {author} {\bibinfo {author} {\bibfnamefont {V.}~\bibnamefont
  {Zlati\ifmmode~\acute{c}\else \'{c}\fi{}}}\ and\ \bibinfo {author}
  {\bibfnamefont {J.~K.}\ \bibnamefont {Freericks}},\ }\Doi
  {10.1103/PhysRevLett.109.266601} {\bibfield  {journal} {\bibinfo  {journal}
  {Phys. Rev. Lett.},\ }\textbf {\bibinfo {volume} {109}},\ \bibinfo {pages}
  {266601} (\bibinfo {year} {2012})}\BibitemShut {NoStop}%
\bibitem [{\citenamefont {Haldane}(1978)}]{Haldane:prl1978}%
  \BibitemOpen
  \bibfield  {author} {\bibinfo {author} {\bibfnamefont {F.~D.~M.}\
  \bibnamefont {Haldane}},\ }\href
  {http://link.aps.org/doi/10.1103/PhysRevLett.40.416} {\bibfield  {journal}
  {\bibinfo  {journal} {Physical Review Letters},\ }\textbf {\bibinfo {volume}
  {40}},\ \bibinfo {pages} {416} (\bibinfo {year} {1978})}\BibitemShut
  {NoStop}%
\bibitem [{\citenamefont {Okiji}\ and\ \citenamefont
  {Kawakami}(1983)}]{Okiji:prl1983}%
  \BibitemOpen
  \bibfield  {author} {\bibinfo {author} {\bibfnamefont {A.}~\bibnamefont
  {Okiji}}\ and\ \bibinfo {author} {\bibfnamefont {N.}~\bibnamefont
  {Kawakami}},\ }\href {http://link.aps.org/doi/10.1103/PhysRevLett.50.1157}
  {\bibfield  {journal} {\bibinfo  {journal} {Physical Review Letters},\
  }\textbf {\bibinfo {volume} {50}},\ \bibinfo {pages} {1157} (\bibinfo {year}
  {1983})}\BibitemShut {NoStop}%
\bibitem [{\citenamefont {Kawakami}\ and\ \citenamefont
  {Okiji}(1983)}]{Kawakami:1983}%
  \BibitemOpen
  \bibfield  {author} {\bibinfo {author} {\bibfnamefont {N.}~\bibnamefont
  {Kawakami}}\ and\ \bibinfo {author} {\bibfnamefont {A.}~\bibnamefont
  {Okiji}},\ }\href {http://link.aps.org/doi/10.1103/PhysRevLett.51.2011}
  {\bibfield  {journal} {\bibinfo  {journal} {Physical Review Letters},\
  }\textbf {\bibinfo {volume} {51}},\ \bibinfo {pages} {2011} (\bibinfo {year}
  {1983})}\BibitemShut {NoStop}%
\bibitem [{\citenamefont {Zlatic}\ \emph {et~al.}(2013)\citenamefont {Zlatic},
  \citenamefont {Boyd},\ and\ \citenamefont {Freericks}}]{Zlatic2013}%
  \BibitemOpen
  \bibfield  {author} {\bibinfo {author} {\bibfnamefont {V.}~\bibnamefont
  {Zlatic}}, \bibinfo {author} {\bibfnamefont {G.~R.}\ \bibnamefont {Boyd}}, \
  and\ \bibinfo {author} {\bibfnamefont {J.~K.}\ \bibnamefont {Freericks}},\
  }\href@noop {} {\bibfield  {journal} {\bibinfo  {journal} {arXiv:1307.4800}}
  (\bibinfo {year} {2013})}\BibitemShut {NoStop}%
\bibitem [{\citenamefont {Pruschke}\ \emph {et~al.}(1995)\citenamefont
  {Pruschke}, \citenamefont {Jarrell},\ and\ \citenamefont
  {Freericks}}]{Pruschke:AdPhys1995}%
  \BibitemOpen
  \bibfield  {author} {\bibinfo {author} {\bibfnamefont {T.}~\bibnamefont
  {Pruschke}}, \bibinfo {author} {\bibfnamefont {M.}~\bibnamefont {Jarrell}}, \
  and\ \bibinfo {author} {\bibfnamefont {J.~K.}\ \bibnamefont {Freericks}},\
  }\bibfield  {booktitle} {\emph {\bibinfo {booktitle} {Advances in Physics}},\
  }\Doi {10.1080/00018739500101526} {\bibfield  {journal} {\bibinfo  {journal}
  {Advances in Physics},\ }\textbf {\bibinfo {volume} {44}},\ \bibinfo {pages}
  {187} (\bibinfo {year} {1995})}\BibitemShut {NoStop}%
\bibitem [{\citenamefont {Peterson}\ and\ \citenamefont
  {Shastry}(2010)}]{Peterson:prb10}%
  \BibitemOpen
  \bibfield  {author} {\bibinfo {author} {\bibfnamefont {M.~R.}\ \bibnamefont
  {Peterson}}\ and\ \bibinfo {author} {\bibfnamefont {B.~S.}\ \bibnamefont
  {Shastry}},\ }\href {http://link.aps.org/doi/10.1103/PhysRevB.82.195105}
  {\bibfield  {journal} {\bibinfo  {journal} {Physical Review B},\ }\textbf
  {\bibinfo {volume} {82}},\ \bibinfo {pages} {195105} (\bibinfo {year}
  {2010})}\BibitemShut {NoStop}%
\end{thebibliography}
%

\end{document}